\documentclass[fleqn,usenatbib]{mnras}

\usepackage{newtxtext,newtxmath}

\usepackage[T1]{fontenc}
\usepackage{ae,aecompl}

\usepackage{graphicx}	
\usepackage{amsmath}	
\usepackage{amssymb}	
\usepackage{hhline}
\usepackage{multirow}
\usepackage{lineno}
\usepackage{eso-pic}
\usepackage{booktabs}
\title[C{\small IV} Black Hole Masses with OzDES]{C{\Large IV} Black Hole Mass Measurements with the Australian Dark Energy Survey (OzDES)}

\author[J.K. Hoormann et al.]{
\parbox{\textwidth}{
\Large
J.~K.~Hoormann,$^{1}$
P.~Martini,$^{2,3}$
T.~M.~Davis,$^{1}$
A.~King,$^{4}$
C.~Lidman,$^{5}$
D.~Mudd,$^{6}$
R.~Sharp,$^{5}$
N.~E.~Sommer,$^{5}$
B.~E.~Tucker,$^{5}$
Z.~Yu,$^{3}$
S.~Allam,$^{7}$
J.~Asorey,$^{8}$
S.~Avila,$^{9}$
M.~Banerji,$^{10,11}$
D.~Brooks,$^{12}$
E.~Buckley-Geer,$^{7}$
D.~L.~Burke,$^{13,14}$
J.~Calcino,$^{1}$
A.~Carnero~Rosell,$^{15,16}$
D.~Carollo,$^{17}$
M.~Carrasco~Kind,$^{18,19}$
J.~Carretero,$^{20}$
F.~J.~Castander,$^{21,22}$
M.~Childress,$^{23}$
J.~De~Vicente,$^{15}$
S.~Desai,$^{24}$
H.~T.~Diehl,$^{7}$
P.~Doel,$^{12}$
B.~Flaugher,$^{7}$
P.~Fosalba,$^{21,22}$
J.~Frieman,$^{7,25}$
J.~Garc\'ia-Bellido,$^{26}$
D.~W.~Gerdes,$^{27,28}$
D.~Gruen,$^{29,13,14}$
G.~Gutierrez,$^{7}$
W.~G.~Hartley,$^{12,30}$
S.~R.~Hinton,$^{1}$
D.~L.~Hollowood,$^{31}$
K.~Honscheid,$^{2,32}$
B.~Hoyle,$^{33,34}$
D.~J.~James,$^{35}$
E.~Krause,$^{36}$
K.~Kuehn,$^{37}$
N.~Kuropatkin,$^{7}$
G.~F.~Lewis,$^{38}$
M.~Lima,$^{39,16}$
E.~Macaulay,$^{9}$
M.~A.~G.~Maia,$^{16,40}$
F.~Menanteau,$^{18,19}$
C.~J.~Miller,$^{27,28}$
R.~Miquel,$^{41,20}$
A.~M\"oller,$^{5}$
A.~A.~Plazas,$^{42}$
A.~K.~Romer,$^{43}$
A.~Roodman,$^{13,14}$
E.~Sanchez,$^{15}$
V.~Scarpine,$^{7}$
M.~Schubnell,$^{28}$
S.~Serrano,$^{21,22}$
I.~Sevilla-Noarbe,$^{15}$
M.~Smith,$^{23}$
R.~C.~Smith,$^{44}$
M.~Soares-Santos,$^{45}$
F.~Sobreira,$^{46,16}$
E.~Suchyta,$^{47}$
E.~Swann,$^{9}$
M.~E.~C.~Swanson,$^{19}$
G.~Tarle,$^{28}$
S.~A.~Uddin,$^{48}$
and (DES Collaboration)
\\ 
{\it \footnotesize Affiliations are listed at the end of the paper}
}
}

\AddToShipoutPictureBG*{%
  \AtPageUpperLeft{%
    \hspace{0.75\paperwidth}%
    \raisebox{-3.5\baselineskip}{%
      \makebox[0pt][l]{\textnormal{DES-2018-0405}}
}}}%

\AddToShipoutPictureBG*{%
  \AtPageUpperLeft{%
    \hspace{0.75\paperwidth}%
    \raisebox{-4.5\baselineskip}{%
      \makebox[0pt][l]{\textnormal{FERMILAB-PUB-19-018-AE}}
}}}%
\date{Accepted XXX. Received YYY; in original form ZZZ}

\pubyear{2019}

\begin{document}

\label{firstpage}
\pagerange{\pageref{firstpage}--\pageref{lastpage}}
\maketitle

\begin{abstract}
Black hole mass measurements outside the local universe are critically important to derive the growth of supermassive black holes over cosmic time, and to study the interplay between black hole growth and galaxy evolution. In this paper we present two measurements of supermassive black hole masses from reverberation mapping (RM) of the broad \ion{C}{iv} emission line. These measurements are based on multi-year photometry and spectroscopy from the Dark Energy Survey Supernova Program (DES-SN) and the Australian Dark Energy Survey (OzDES), which together constitute the OzDES RM Program. The observed reverberation lag between the DES continuum photometry and the OzDES emission-line fluxes is measured to be $358^{+126}_{-123}$ and $343^{+58}_{-84}$ days for two quasars at redshifts of $1.905$ and $2.593$ respectively. The corresponding masses of the two supermassive black holes are $4.4 \times 10^{9}$ and $3.3 \times 10^{9}$ M$_\odot$, which are among the highest-redshift and highest-mass black holes measured to date with RM studies. We use these new measurements to better determine the \ion{C}{iv} radius$-$luminosity relationship for high-luminosity quasars, which is fundamental to many quasar black hole mass estimates and demographic studies. 

\end{abstract}




\section{Introduction}

The masses of the supermassive black holes at the centres of galaxies are fundamental measurements for studies of galaxy evolution, active galactic nuclei (AGN), and the interaction between black hole and galaxy growth over cosmic time. Mass measurements in the local universe are most often obtained with high-spatial resolution spectroscopy that can resolve the sphere of influence of the black hole \citep{Kormendy1995,Richstone1998,Ferrarese2005, Kormendy2013}, yet even the largest telescopes lack sufficient angular resolution for such measurements at significant distances.

Fortunately such measurements are possible for AGN with broad emission lines, irrespective of distance, through intensive time-domain spectrophotometry. The technique of reverberation mapping (RM) resolves very small scales with measurement of the time lag between variations in the continuum and the broad line region \citep{Blandford1982,Peterson1993}. This provides a measurement of the distance of the broad emission-line region (BLR) from the supermassive black hole because the continuum variations originate in the accretion disks on scales of only a few Schwarzschild radii, while the BLR is approximately an order of magnitude or more further away. The time lag, $\tau$, between the continuum variation and when the BLR reverberates in response is consequently a measurement of the size of the BLR: $R_\mathrm{BLR} = c\tau$. This measurement is combined with a measurement of the characteristic velocity, $\Delta V$, of the BLR as measured by spectral line widths to determine the black hole mass through an application of the virial theorem:

\begin{equation}
M_\mathrm{\rm{BH}} = f \frac{R_\mathrm{BLR} \Delta V^2}{G}. \label{eq:bhmass}
\end{equation}

\noindent The quantity $f$ is a dimensionless factor that accounts for the geometry, orientation, and kinematics of the BLR. 

The factor $f$ has been measured with reverberation-based black hole mass measurements for many nearby galaxies that also have black hole mass estimates from the correlation between  black hole mass and the stellar velocity dispersion of the host galaxy spheroid known as the $M_\mathrm{\rm{BH}} - \sigma_*$ relationship \citep{Ferrarese2000,Gebhardt2000}. This relationship has been measured for large numbers of nearby quiescent \citep{Tremaine2002,McConnell2013} and active galaxies \citep{Gebhardt2000b,Woo2010,Grier2013,Grier2017b}. The ensemble average value of $f$ is $\langle f \rangle = 4.47 \pm 1.25$ based on about 30 AGN with both $\sigma_*$ and $M_\mathrm{\rm{BH}}$ measurements from reverberation mapping \citep{Woo2015}. This average value for $f$ is also consistent with measurements of two nearby galaxies that both have reverberation-based masses and measurements from spatially-resolved kinematics \citep{Davies2006,Hicks2008,Onken2014} and sophisticated dynamical models of the BLR \citep{Pancoast2014, Williams2018}.

Most reverberation mapping campaigns have targeted variable AGN in the local universe ($z < 0.3$) to measure the response of the prominent H$\beta$ line \citep{Kaspi2000,Peterson2004,Bentz2009,Denney2010,Grier2012}. These measurements showed that there is a tight relationship  between $R_\mathrm{BLR}$ and the AGN luminosity, the $R-L$ relationship \citep{Bentz2009}. The power-law slope of the relationship is $\alpha = 0.533^{+0.035}_{-0.033}$, which is consistent with the value of $0.5$ expected from simple photoionization arguments, and the scatter around this slope is only $0.13 \pm 0.02$ dex for the best subset of the data \citep{Bentz2013}. The $R-L$ relationship is extraordinarily useful because it provides a way to estimate the black hole mass of an AGN with a single spectroscopic measurement of the H$\beta$ region, rather than the many tens to even a hundred epochs required to measure a reverberation-based mass for a single object \citep{Wandel1999,Vestergaard2002,Vestergaard2006}. The small scatter also led to the suggestion that the $R-L$ could be used to treat AGN as standard candles \citep{Watson2011}. The visibility of luminous AGN out to higher redshifts than type Ia supernovae make them attractive probes of some dark energy models \citep{King2014}. 

At higher redshifts, it is either not possible or much more challenging to measure the H$\beta$ emission line. Various studies have consequently used spectroscopic observations of select samples of quasars to produce empirical methods to estimate black hole masses with other broad emission lines, most notably the prominent \ion{Mg}{ii} and \ion{C}{iv} lines \citep{McLure2002,Vestergaard2006,Wang2009}. These methods typically require extrapolations to higher-redshift and higher-luminosity AGN, and are also more uncertain due to differences in the geometry and kinematics of the \ion{Mg}{ii} and \ion{C}{iv} line regions relative to H$\beta$. This has consequently inspired efforts to directly measure reverberation masses for high-redshift AGN with these emission lines. These studies have measured masses for 19 AGN with the \ion{C}{iv} line \citep{Peterson2004b,Peterson2005,Metzroth2006,Kaspi2007,Trevese2014,Lira2018} and 10 with the \ion{Mg}{ii} line \citep{Clavel1991,Reichert1994,Metzroth2006,Shen2015}, in addition to reverberation mapping of higher-redshift AGN with the H$\beta$ line \citep{Shen2016,Grier2017}.  These measurements indicate that H$\beta$ and \ion{Mg}{ii} emanate at the same radius with \ion{C}{iv} originating from the inner regions of the BLR clouds \citep{Kaspi2007, Trakhtenbrot2012}.  However, only a few AGN currently have lags measured using multiple emission lines which are necessary to fully understand the stratification of the BLR.

In this paper we present some of the first results from the OzDES Reverberation Mapping (OzDES RM) Program \citep{King2015}, which combines spectroscopic observations from the Australian Dark Energy Survey \citep[OzDES, see][]{Yuan2015,Childress2017} with photometric data from the Dark Energy Survey \citep[DES, see][]{Flaugher2005, DES2016}. Section \ref{sec:Observations} details the observations obtained by OzDES and the data calibration procedures. Section \ref{sec:Measurements} describes our measurements of the emission line fluxes and reverberation lags, and the resultant black hole masses for two high-redshift AGN. We use these new measurements together with other measurements in the literature to calculate a new \ion{C}{iv} $R-L$ relationship in Section \ref{sec:Results} and discuss and summarise our results and the outlook to the future in Section \ref{sec:Discussion}.

\section{Observations} \label{sec:Observations}

The OzDES RM Program combines spectroscopic observations from the OzDES survey and photometric observations from DES-SN. DES is conducted with the CTIO 4m Blanco Telescope using the Dark Energy Camera \citep[DECam][]{Flaugher2015} in the $grizY$ bands. The DES footprint comprises 5000 square degrees at high Galactic latitude that are visible from the southern hemisphere \citep{Diehl2016,Diehl2018}. The main science goal of DES is to study the expansion of the universe through four cosmological probes: weak gravitational lensing \citep{DESWL}, galaxy clustering \citep{DESGC}, baryon acoustic oscillations \citep{DESBAO}, and supernovae (SN). The first three cosmological probes use data from the wide-area survey. DES-SN uses approximately weekly observations of ten dedicated SN fields that together comprise approximately 27 square degrees \citep{Kessler2015, DAndrea2018}. After Science Verification in the 2012 semester, DES began in 2013 and completed the wide-area survey in January 2019 (the supernova survey ended in 2017).

OzDES is a spectroscopic follow-up program for the 10 DES-SN fields. The survey is carried out using the AAOmega spectrograph \citep{Smith2004} and the Two Degree Field (2dF) 400 multi object fibre positioning system \citep{Lewis2002}, which covers the wavelength range of 3700-8800{\AA}. The primary goal of OzDES is to obtain redshifts for SN host galaxies \citep{Yuan2015,Childress2017}.  OzDES was awarded 100 nights on the 3.9m Anglo-Australian Telescope (AAT) that were originally planned over five years. Twelve nights were reallocated from the fourth and fifth season into a sixth season in 2018B. The purpose of this reallocation was to allow time to measure SN host galaxy redshifts for discoveries that were made at the end of 2017B.  

\subsection{OzDES Reverberation Mapping Candidates}

In addition to supernova host galaxies, the OzDES survey also targets AGN for the OzDES RM program, as well as a number of other ancillary target classes \citep{Childress2017}. The possible AGN targets in the DES-SN fields were selected based on colour and variability using the data from the DES Year 1 catalogues, DES Y1A1.  These targets were spectroscopically confirmed by OzDES \citep{Tie2017}.  The OzDES RM sample was chosen based on the presence of clear H$\beta$, \ion{Mg}{ii}, and/or \ion{C}{iv} lines with a median signal to noise ratio (SNR) in the line greater than 10 as measured from spectra of a larger sample of AGN that were observed during the first two years of OzDES.  To enhance the science output, a lower SNR$>5$ threshold was used for AGN at $z>3$ and/or for AGN with more than one broad emission line in the spectral bandpass \citep{King2015}. Additionally, AGN that had fewer than 10 epochs of photometric observations after the first DES season were removed from the sample. Such AGN typically were close to or in the DECam CCD gaps \citep{Tie2017, Mudd2017b}. The final OzDES RM sample includes 771 AGN with $0.1 < z < 4.5$. 

An extensive simulation study by \citet{King2015} shows that OzDES will be able to measure time lags for $30-40$\% of the AGN once spectroscopic epochs are observed for all six seasons. This success rate is highly-dependent on redshift and other factors, such as SNR and the calibration uncertainty. For example, the success rate for the \ion{C}{iv} line is expected to be closer to 20-30\% \citep{King2015}. At present the first 4 years of DES photometry have been processed through the DES Data Management System and are available for analysis \citep{Morganson2018}, as well as for calibration of the OzDES spectroscopy . On average these data include about 110 photometric epochs and 15 spectroscopic epochs.  Most AGN have four additional spectroscopic epochs that were obtained during the fifth year of DES, and four additional OzDES observing runs have been completed for Year 6.

\subsection{Spectroscopic Calibration} \label{sec:Calibration}

\begin{figure}
\includegraphics[width=\columnwidth]{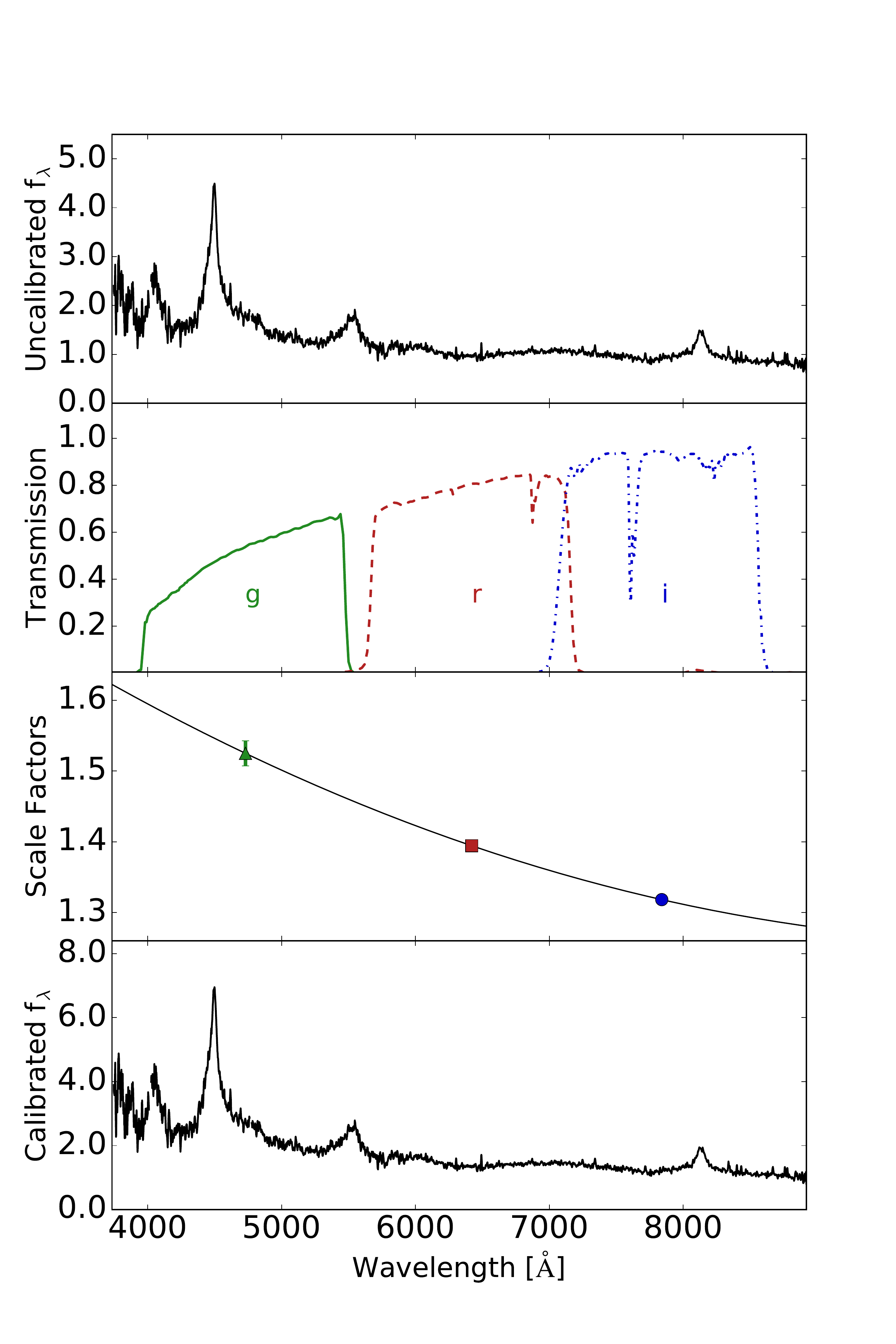}
\caption{Single smoothed spectrum of the AGN DES J022828.19$-$040044.30 taken on MJD 56917.615 ({\it top panel}). We calculate OzDES instrumental magnitudes in the $gri$ bands with the DES transmission functions ({\it second panel}). The scale factors are in units of $10^{-16}$~ergs~s$^{-1}$~cm$^{-2}${\AA}$^{-1}$~counts$^{-1}$. We fit a second-order polynominal to these scale factors and use this function to convert the OzDES spectroscopic pipeline output to the fluxes measured from contemporaneous DES photometry ({\it third panel}). The calibrated spectrum is shown in the bottom panel. The flux is in units of $10^{-16}$~erg~s$^{-1}$~cm$^{-2}${\AA}$^{-1}$. \label{fig:scaleFactors}}
\end{figure}

\begin{figure}
        \includegraphics[width=\columnwidth]{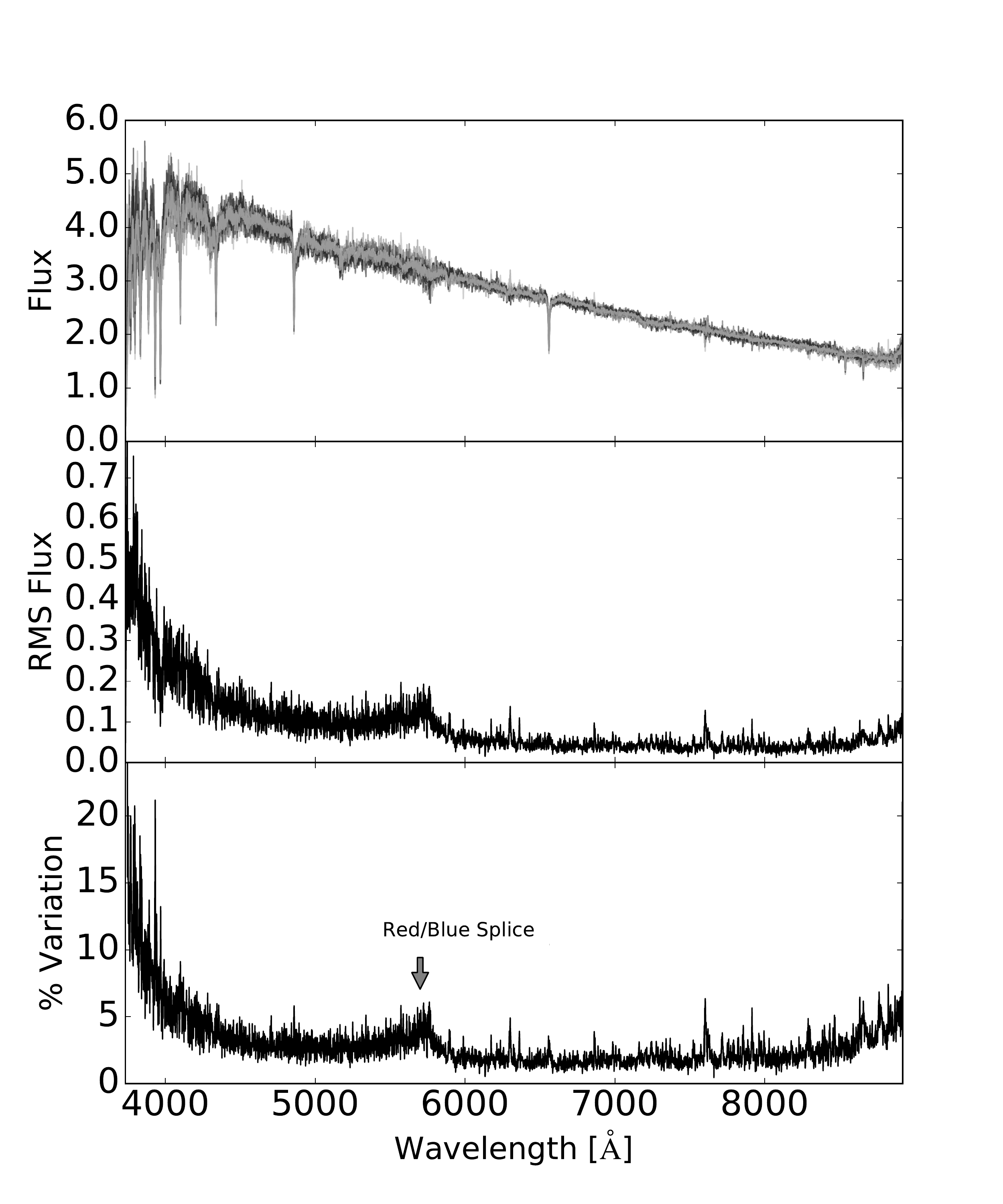}
    \caption{Seventeen spectra of the F star FSC0225$-$0444 taken throughout OzDES operations which we spectrophotometrically calibrated ({\it top panel}). In the middle panel we show the RMS spectrum of these calibrated observations. We found the average variation ({\it bottom panel}) to be less than 5\% of the mean flux. The bump at 5700{\AA} is due to the dichrotic split between the red and the blue arm of the spectrograph with the noise at the far blue end being due to the reduced count rate and modest transmission losses in the flat field lamps. All fluxes are in units of $10^{-16}$~ergs~s$^{-1}$~cm$^{-2}${\AA}$^{-1}$. \label{fig:fStar}}

\end{figure}


The high-quality DES photometry calibration provides good measurements of the continuum flux variations of all of the AGN in our sample \citep{Burke2018}. Emission-line reverberation mapping also requires measurements of the emission-line flux variations from the OzDES spectra. These spectra are obtained with a fibre spectrograph, and consequently the amount of flux in the fibre at each epoch depends on many factors, most notably the image quality, airmass, transparency, and the accuracy of the fibre placement.  Furthermore, these factors can impact not just the total flux that enters the fibre at each epoch, but also the wavelength-dependent flux calibration for each epoch. Two wavelength-dependent factors that are important are the variation in image quality (seeing) with wavelength and chromatic effects in the spectrograph optics, where the latter could depend on location in the field of view.  A common practice with fibre spectroscopy is to flux-calibrate the spectra with photometric measurements in multiple, broad-band filters. For example, \citet{Hopkins2013} used SDSS photometry to flux-calibrate spectra from the AAOmega spectrograph for the Galaxy And Mass Assembly (GAMA) sample and obtained a typical flux calibration uncertainty of about 10\%, although the quality degraded to somewhat poorer than 20\% at the extreme ends of the wavelength range.

The variability of AGN precludes the use of a single photometric measurement for flux calibration. Instead we take advantage of the near-weekly cadence of the DES-SN observations, which provide DES photometric measurements in the $griz$ filters within a week of the OzDES epochs\footnote{Occasionally particularly bad or good weather has led to observations of the same fields over multiple nights in a single OzDES run. In these cases we combine the multiple nights of observations into a single OzDES epoch.}. We linearly interpolate between the photometric epochs immediately before and after each spectroscopic epoch to estimate the fluxes in the $gri$ filters at our spectroscopic epoch. These three bands overlap nearly perfectly with the spectroscopic bandpass. Next, we compute instrumental fluxes in these three bands from the OzDES spectra, fit a second-order polynomial to the flux ratio at the effective wavelength of each bandpass, and use this polynomial to calibrate each OzDES epoch. Figure~\ref{fig:scaleFactors} shows an example of our approach. The top panel shows a single spectroscopic epoch which was smoothed using an exponential smoothing kernel with a decay factor of 0.9 and search window of 11 pixels.  The second panel shows the $gri$ photometric bands that overlap the OzDES wavelength range, the third panel shows the ratio of the instrumental and DES flux values, as well as the polynomial fit, and the bottom panel shows the AGN spectrum after the flux calibration. Through the use of Gaussian processes use the measured uncertainties in the scale factors which were dominated by the photometric uncertainty, to determine the wavelength-dependent variance on the calibration model. This variance was added in quadrature with the other sources of uncertainty.  This spectrophotometric code is publicly available. \footnote{\url{https://github.com/jhoormann/OzDES_calibSpec}}  The calibration is performed on each individual observation before the data is combined.  By coadding the data taken during a given observing run the noise of the spectrum is reduced \citet{Childress2017}.

The quality of the spectroscopic flux calibration is critical to our emission-line reverberation mapping program. As noted in the simulation study of \citet{King2015}, using the flux calibration uncertainties measured by GAMA would result in many of our emission-line light curves being limited more by flux calibration than the signal-to noise ratio of the spectroscopy. We consequently have made numerous improvements to the calibration protocols and software pipeline in order to improve the flux calibration of the OzDES spectra taken at the AAT. These include upgrades to the detectors and additional dome flat field calibrations that better account for relative wavelength-dependent transmission between fibres. These improvements are described in \citet{Yuan2015} for Year 1 of OzDES and in \citet{Childress2017} for changes through the end of Year 3.

An important part of our calibration strategy is observations of tens of F stars in each field. These F star observations have provided a valuable check on improvements to the calibration and data processing procedures, as well as a convenient way to quantify the calibration uncertainties.  Figure~\ref{fig:fStar} demonstrates one measurement of the calibration uncertainty from the F star FSC0225$-$0444. The top panel shows 17 spectra of the star that have been individually calibrated from the DES photometry.  As the F stars do not vary significantly, we use only the mean $gri$ fluxes, measured from the 2012B semester,  to compute the scale factors for each epoch. The middle spectrum shows the RMS flux computed from the 17 calibrated spectra, and the bottom spectrum shows the percent variation. This variation spectrum demonstrates that the flux calibration is better than 5\% over most of the observed spectral range, and only is as poor as 10\% at the bluest wavelengths. This significant improvement relative to GAMA is a testament to the many improvements in the instrument, calibration procedures, and software pipeline.

\section{Time Lag Measurements} \label{sec:Measurements}

\begin{figure*}
	\centering
    \includegraphics[width=0.49\linewidth]{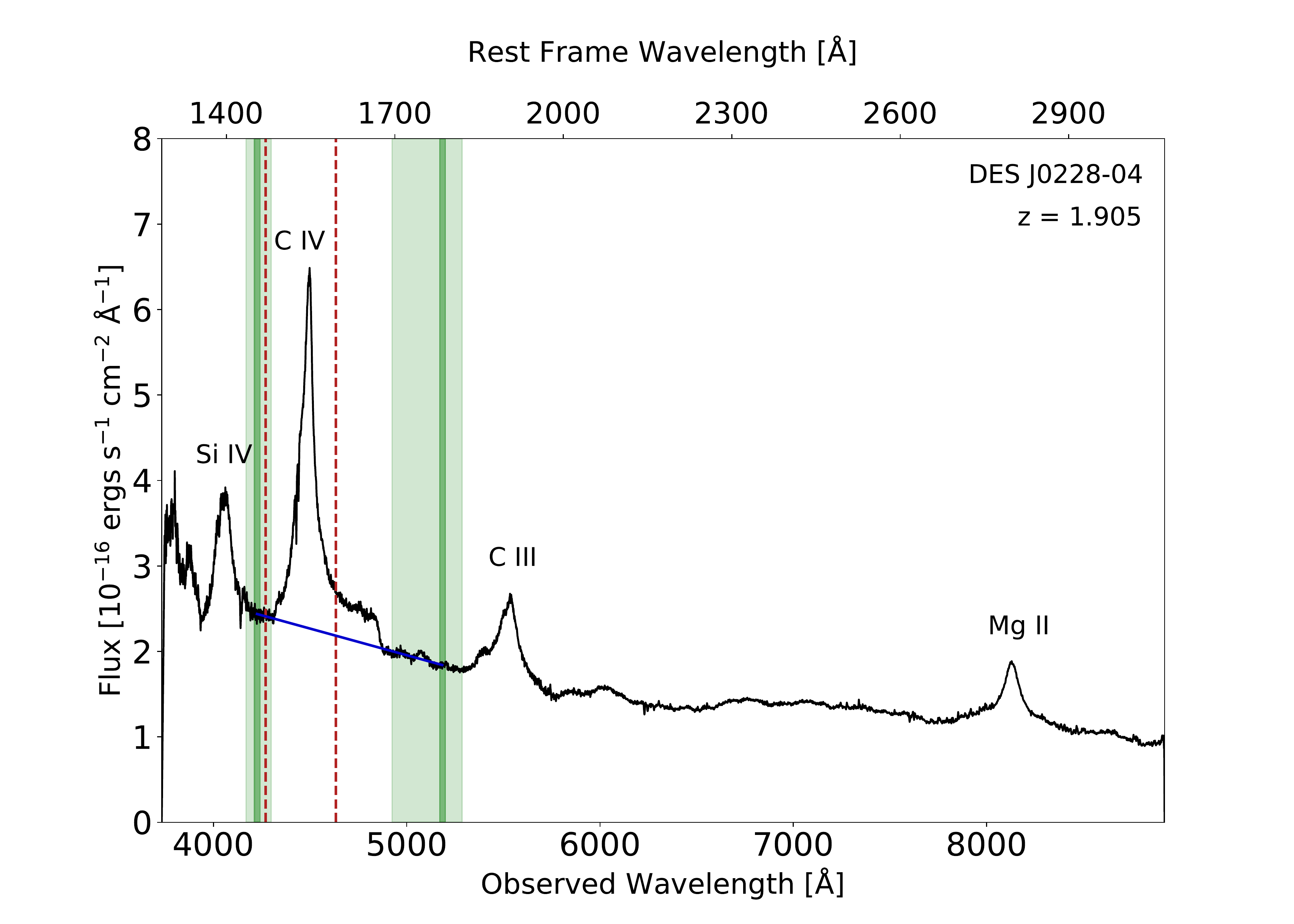}
    \includegraphics[width=0.49\linewidth]{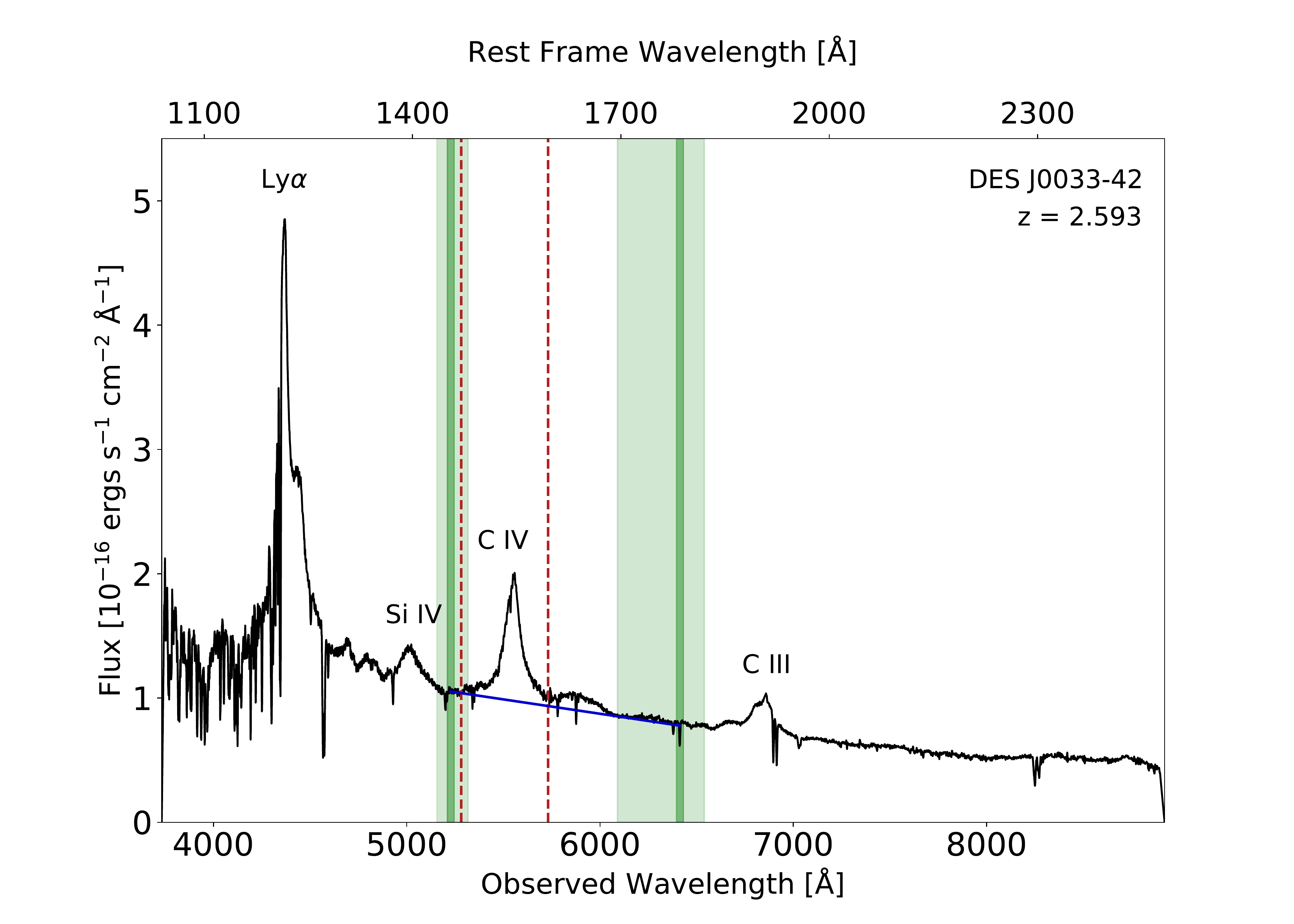}
    \caption{Coadded spectra for both AGN studied in this paper using the first four years of OzDES observations after spectrophotometric calibration. Red dashed lines indicate the integration window chosen for the \ion{C}{iv} line.  The windows used for the continuum subtraction are indicated with dark green and the local continuum model is plotted in blue.  The regions used in the uncertainty calculations are shown in light green. \label{fig:spectra}}
\end{figure*}

\begin{figure*}
	\centering
    \includegraphics[width=0.55\linewidth]{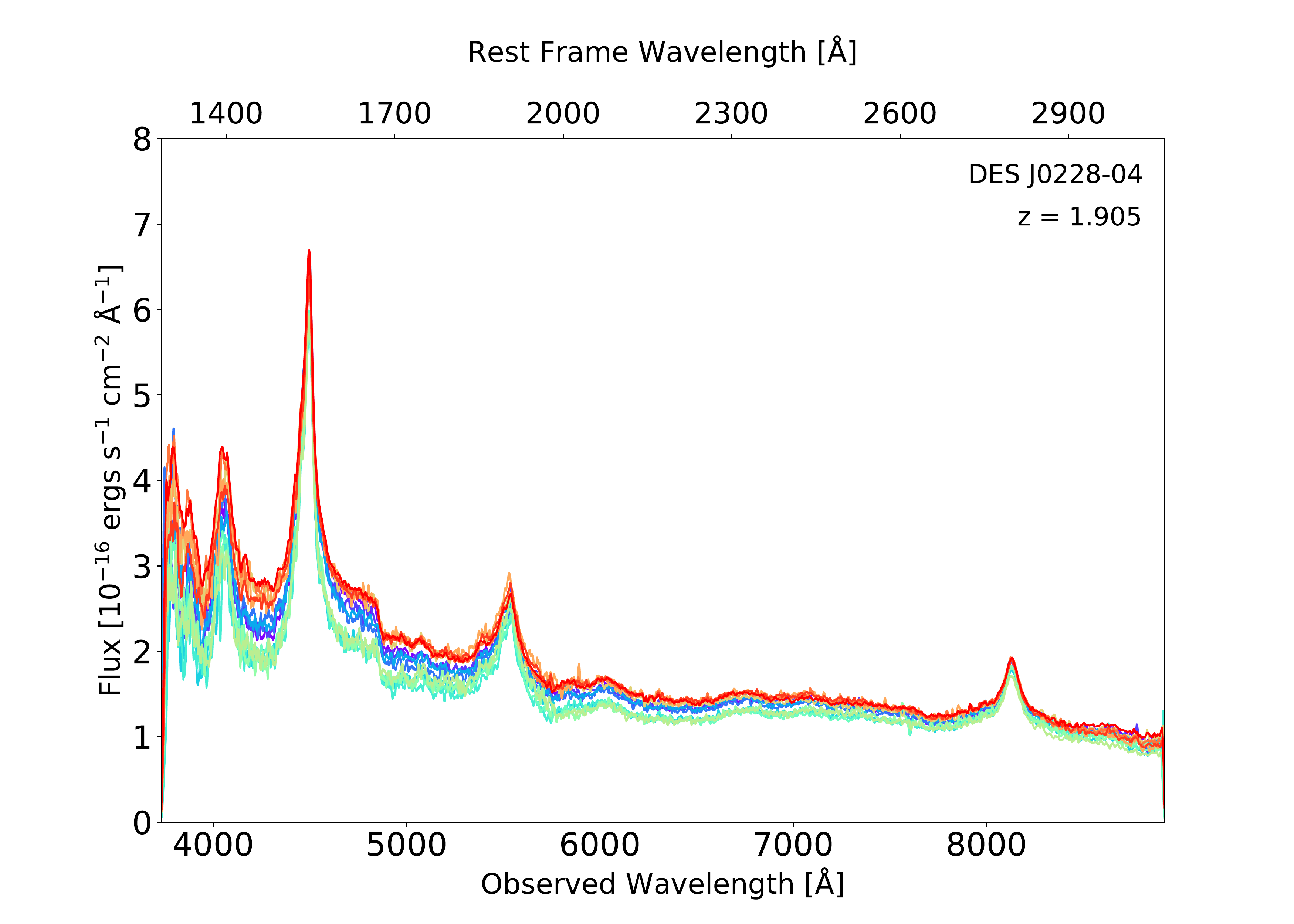}
    \includegraphics[width=0.39\linewidth]{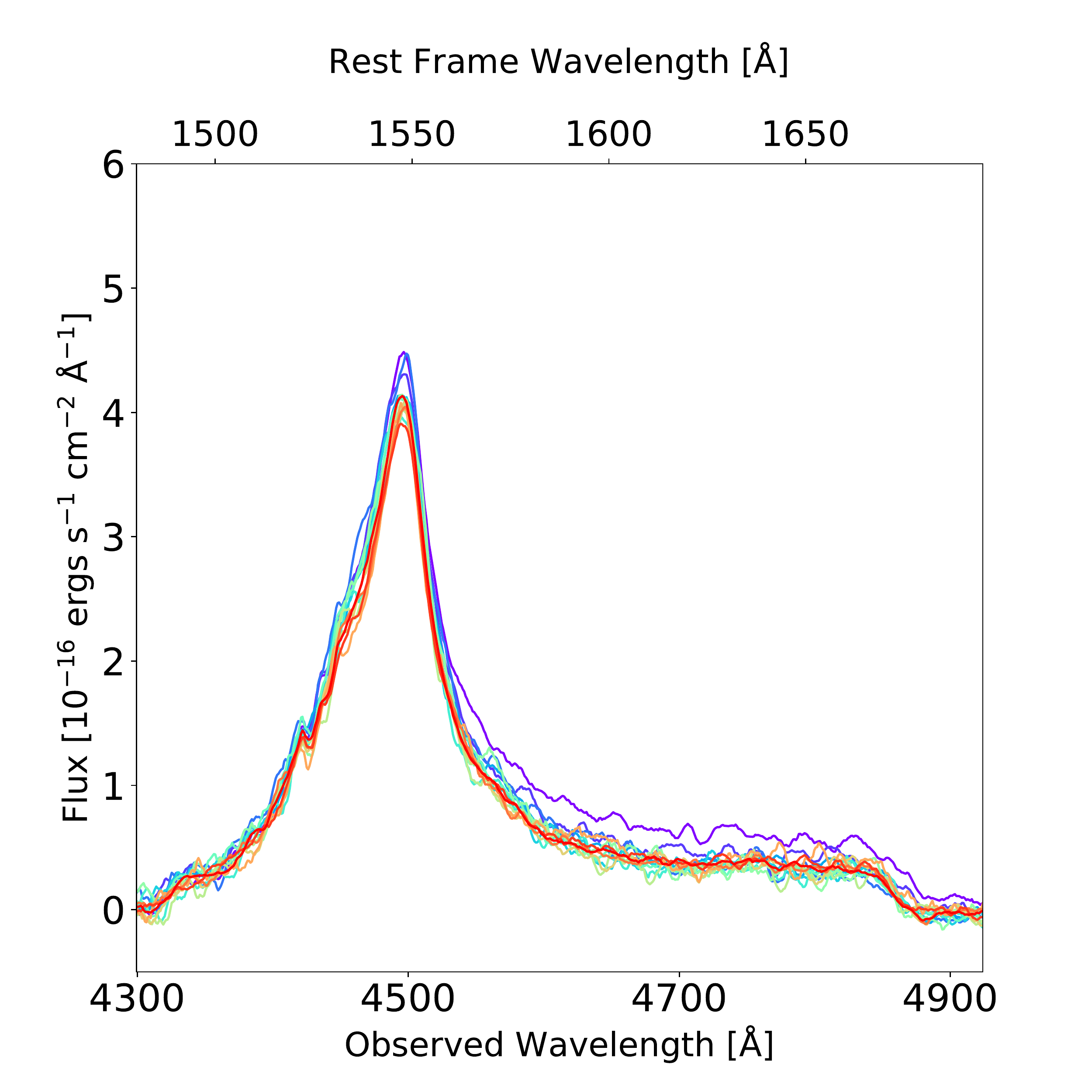}    
    \caption{Smoothed spectrum for each epoch of observation of DES J0228-04. The {\it left panel} shows the full spectrum for each epoch and the {\it right panel} focuses in on the continuum subtracted \ion{C}{iv} line. \label{fig:spectraMany955}}
\end{figure*}

\begin{figure*}
	\centering
    \includegraphics[width=0.55\linewidth]{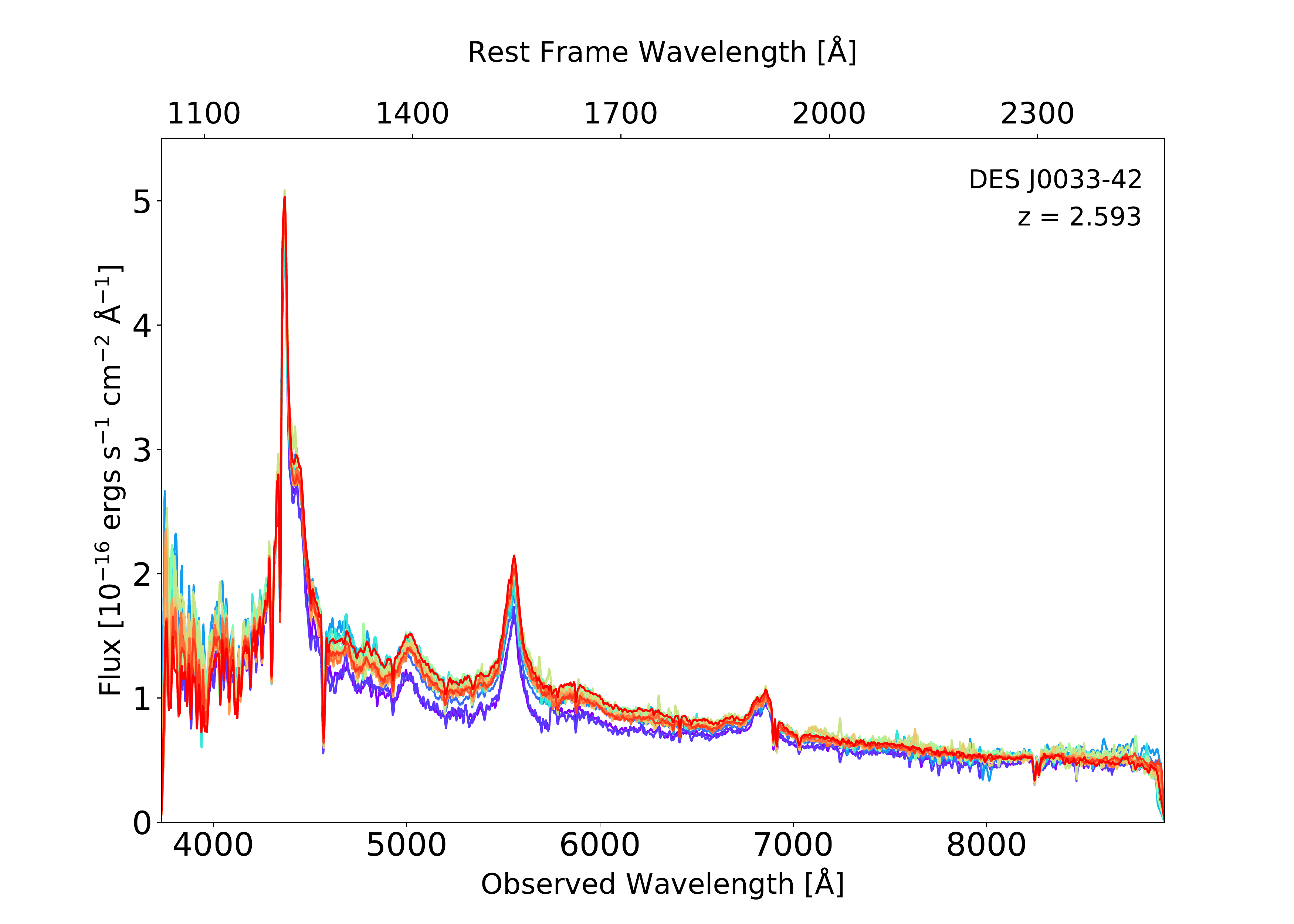}
    \includegraphics[width=0.39\linewidth]{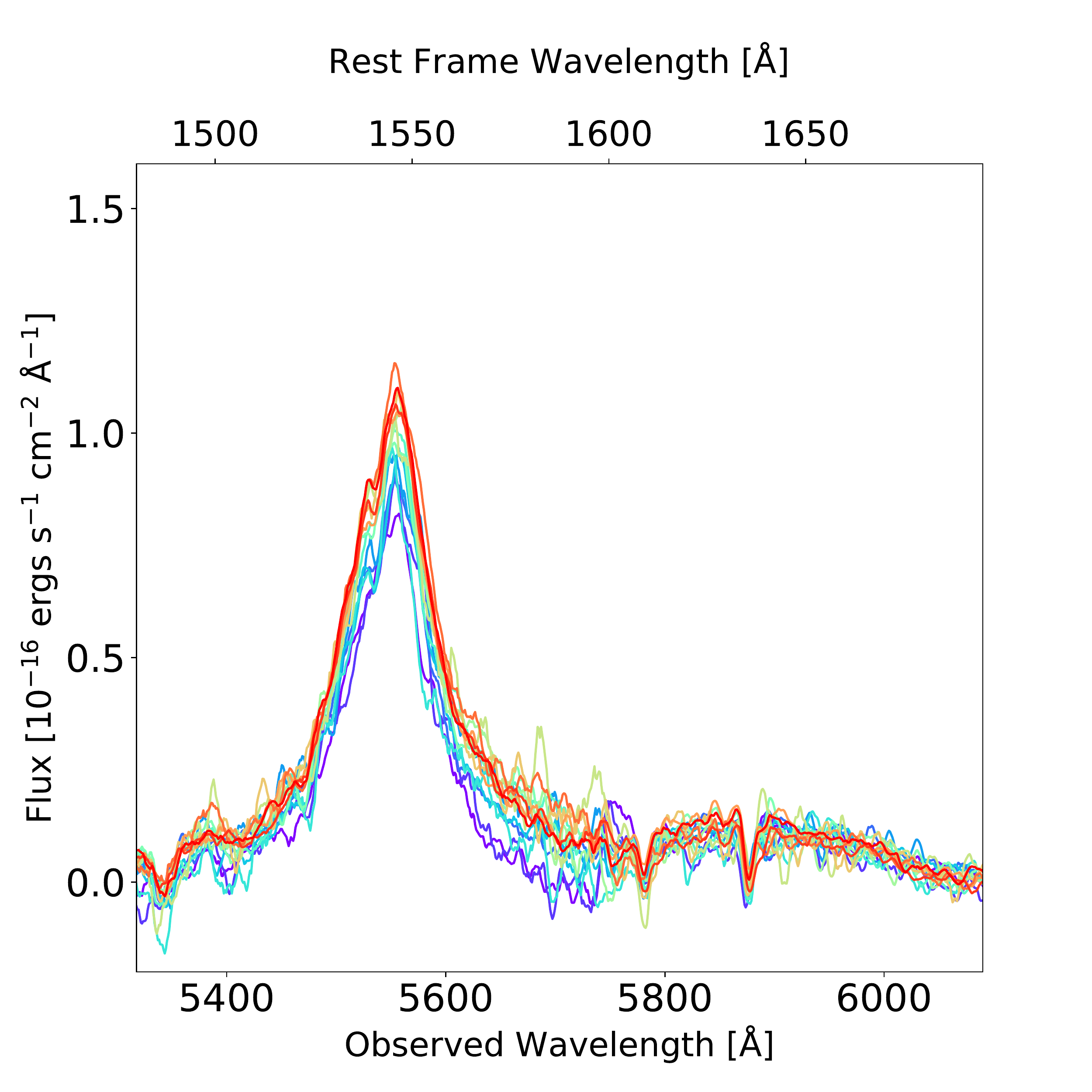}
    \caption{Smoothed spectrum for each epoch of observation of DES J0033-42. The {\it left panel} shows the full spectrum for each epoch and the {\it right panel} focuses in on the continuum subtracted \ion{C}{iv} line. \label{fig:spectraMany466}}
\end{figure*}

Of the 771 AGN OzDES is regularly monitoring,  393 of them are at the requisite redshift for the \ion{C}{iv} line and nearby continuum to be within the wavelength range for OzDES to observe.  We only consider RM using the \ion{C}{iv} line in this paper as contamination from \ion{Fe}{II} emission is not an issue in this wavelength range.  We will present RM results for the \ion{Mg}{ii} and H$\beta$ emission lines in future work.  For this first study we identified a subset of these 393 AGN that were expected to have lags on the order of one year based on the $R-L$ relationship from \citet{Kaspi2007} which would be most easily observed with this partial OzDES data set.  The list was further culled, primarily through visual inspection, to only include AGN with light curves that were variable and had a high cadence.  The subset consisted of 23 AGN.  For this preliminary study using the partial data set, we recovered \ion{C}{iv} lags for two of the AGN in this subset.  The lags were measured for DES J022828.19-040044.30 at $z$ = 1.905  and DES J003352.72-425452.60 at $z$ = 2.593 (DES J0228-04 and DES J0033-42 respectively in future). Figure \ref{fig:spectra} shows the coadded spectra for the first four years of OzDES for both sources.  The spectrum for each individual epoch is shown in the left panel of Figure \ref{fig:spectraMany955} and \ref{fig:spectraMany466} for DES J0228-04 and DES J0033-42.  The right panel isolates the continuum subtracted \ion{C}{iv} line.

\subsection{Line Flux Measurements} \label{sec:Fluxes}

To measure line fluxes, we implemented a local continuum subtraction method to isolate the line variation from the continuum.  We used two regions, located on either side of the \ion{C}{iv} line and free of other prominent emission lines,  to represent the continuum.  We chose these regions, from 1450-1460{\AA} on the blue side and 1780-1790{\AA} on the red side in the rest frame, using the SDSS composite AGN spectrum \citep{VandenBerk2001}.  These regions are indicated in dark green in Figure \ref{fig:spectra}. After calculating the mean wavelength and flux in each continuum region we performed a linear fit to the data in order to model the continuum which we subtract.  The local continuum model for the coadded spectra is given by the blue lines in Figure \ref{fig:spectra}.

The regions chosen to represent the continuum have the potential to significantly impact the overall line flux measurements.  In order to quantify the effect the choice of continuum regions had on the resulting light curves we varied the bounds of these regions and found the resulting line flux.   To do this, we first identified the regions surrounding the \ion{C}{iv} line that  could reasonably be classified as clean enough to represent the continuum, shown by the light green regions in Figure \ref{fig:spectra}. We then performed a bootstrapping procedure by randomly picking subsets of these regions to represent the continuum while resampling the fluxes from a Gaussian distribution defined by their uncertainty. The uncertainty for each flux is determined by the variance spectrum corresponding to the flux spectrum which includes both observational and calibration uncertainties for all observations in the observing run.  We used the standard deviation of the distribution in the resulting line fluxes to quantify the total uncertainty included that obtained through continuum subtraction.  While the regions for continuum subtraction were chosen as they are relatively clean this method does allow for any unexpected features, such as absorption lines, to be taken into account in the uncertainties. The average uncertainty in emission line flux for a given epoch was $\sim$ 3.5\% for DES J0228-04 and $\sim$ 7.7\% for DES J0033-42. Approximately 75\% of this uncertainty is due to the choice of region used to represent the continuum, as the slope of the continuum can significantly affect the line flux measurement.  The remainder of the uncertainty is a result of the variance on the flux measurement due to observational and calibration uncertainties.  The properties of these light curves including mean cadence, uncertainty, and excess variance \citep{Rodrguez1997} are shown in Table \ref{tab:sourceSummary}.

\begin{table*}
	\centering
	\caption{Light curve properties for the two AGN in this study including mean cadence and \% error, and number of epochs during the baseline of observation.}
	\label{tab:sourceSummary}
	\begin{tabular}{|lccccccccc}
		\hline
		\multicolumn{2}{l}{ } & \multicolumn{4}{c}{\ion{C}{iv}}  & \multicolumn{4}{c}{g-Band} \\
		AGN & $z$ &  epochs &  cadence [days] &  \% error & $F_{\rm{var}}$ & epochs &  cadence [days] &  \% error  & $F_{\rm{var}}$ \\ \cmidrule(r){1-2} \cmidrule(l){3-6}  \cmidrule(l){7-10}
		DES J0228-04 & 1.905 & 14 & 31 & 3.5 & 0.04 & 107 & 7 & 0.7 & 0.11 \\
		DES J0033-42 & 2.593 & 15 & 31 & 7.7 & 0.13 & 121 & 6 & 0.9 & 0.07\\
		\hline
	\end{tabular}
\end{table*} 

We calculated the mean flux density of the emission line directly from the continuum subtracted emission line spectrum over a wavelength range that was fixed throughout all observations of the source.  We chose a baseline integration window for the \ion{C}{iv} line of 1470-1595{\AA}.  Through visual inspection we verified that this window both fully included the emission line and excluded the red shelf, a broad plateau seen near the red wing of the \ion{C}{iv} line caused by \ion{He}{ii}~$\lambda$1640 and \ion{O}{iii}~$\lambda$1663 and the unidentified but prominent emission at 1600{\AA} \citep{Fine2010,Assef2011}.  The integration window is indicated by the red dashed lines in Figure \ref{fig:spectra}.

\begin{figure*}
	\centering
    \includegraphics[width=0.45\linewidth]{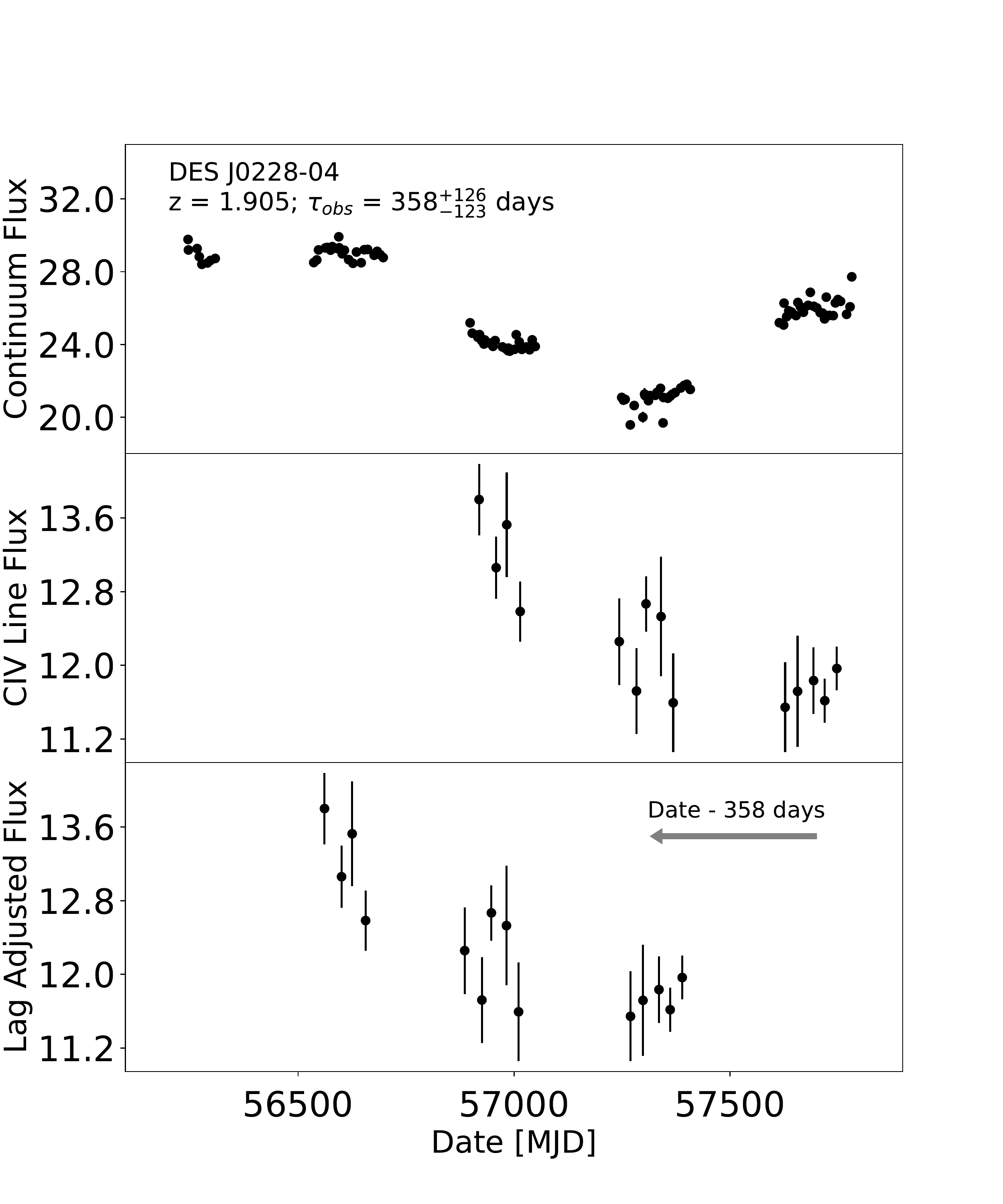}
    \includegraphics[width=0.45\linewidth]{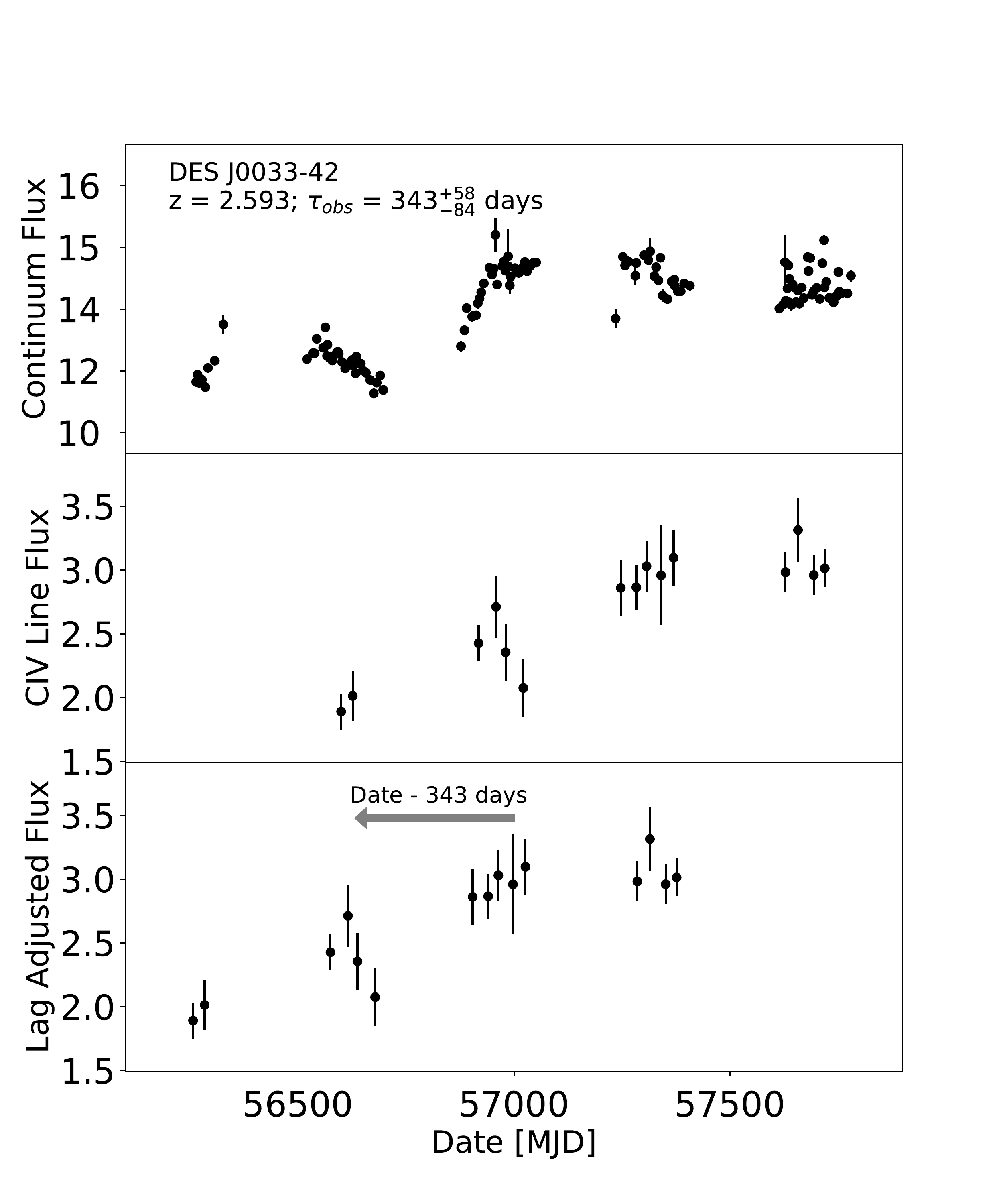}
    \caption{Light curves of the two OzDES AGN in this study including the continuum flux represented by the DES g-band photometry ({\it top panel}) and \ion{C}{iv} line flux measured from the OzDES spectra ({\it middle panel}).  The \ion{C}{iv} line light curve adjusted by the measured time lag is shown in the {\it bottom panel}. Fluxes are in units of $10^{-17}$~ergs~s$^{-1}$~cm$^{-2}${\AA}$^{-1}$. \label{fig:lc}}
\end{figure*}

The light curves for  DES J0228-04 and for DES J0033-42 are shown in Figures \ref{fig:lc}.  The top panel shows the continuum emission, which is from the DES $g$-band photometry. The $g$-band photometry contains the 1350{\AA} wavelength used to represent the luminosity  for high redshift AGN in addition to some variability due to BLR features that occur within the bandpass.  This includes photometry from the first four years of DES plus data from the preceeding Science Verification semester (2012).  The photometric light curves were constructed from the DES Year 4 catalogue, DES Y4A1, and include the DES calibration error from \citet{Burke2018}.  The \ion{C}{iv} line light curves are shown in the middle panel and the \ion{C}{iv} line light curves shifted based on the calculated time lag are shown in the bottom panel.  This includes data from the first four years of OzDES although DES J0228-04 had no data taken during Year 1. The \ion{C}{iv} emission line and photometric light curves are presented in Table \ref{tab:lightcurves}.

\begin{table*}
	\centering
	\caption{\ion{C}{iv} and photometric light curves for the two AGN in this study.  Flux densities are in units of $10^{-17}$ ergs s$^{-1}$ cm$^{-2}${\AA}$^{-1}$. Full light curves for both AGN are available online as machine-readable tables. \label{tab:lightcurves}}
	\begin{tabular}{|lcccccccc}
		\hline
		\multicolumn{1}{l}{ } & \multicolumn{2}{c}{\ion{C}{iv}}  & \multicolumn{2}{c}{g-Band} & \multicolumn{2}{c}{r-Band} & \multicolumn{2}{c}{i-Band} \\
		AGN & MJD & Flux & MJD & Flux & MJD & Flux & MJD & Flux   \\ \cmidrule(r){1-1} \cmidrule(l){2-3}  \cmidrule(l){4-5} \cmidrule(l){6-7} \cmidrule(l){8-9}
DESJ0228-04 & 56919.14 & 13.80 $\pm$ 0.39 & 56245.12 & 29.78 $\pm$ 0.20 & 56245.12 & 16.87 $\pm$ 0.10 & 56253.07 & 14.73 $\pm$ 0.08 \\
 & 56958.57 & 13.06 $\pm$ 0.34 & 56246.07 & 29.20 $\pm$ 0.20 & 56246.07 & 16.69 $\pm$ 0.10 & 56254.11 & 14.51 $\pm$ 0.08 \\
 & 56983.13 & 13.53 $\pm$ 0.57 & 56266.16 & 29.27 $\pm$ 0.20 & 56266.17 & 16.87 $\pm$ 0.10 & 56265.16 & 14.28 $\pm$ 0.08 \\
 & 57014.27 & 12.58 $\pm$ 0.33 & 56271.13 & 28.83 $\pm$ 0.20 & 56271.14 & 16.44 $\pm$ 0.09 & 56271.16 & 14.39 $\pm$ 0.08 \\
 & 57243.76 & 12.26 $\pm$ 0.47 & 56277.16 & 28.41 $\pm$ 0.19 & 56277.17 & 16.30 $\pm$ 0.09 & 56277.14 & 14.34 $\pm$ 0.08 \\
 & 57283.58 & 11.72 $\pm$ 0.47 & 56290.16 & 28.48 $\pm$ 0.20 & 56285.15 & 16.48 $\pm$ 0.10 & 56285.16 & 14.23 $\pm$ 0.08 \\
 & 57305.58 & 12.67 $\pm$ 0.30 & 56297.10 & 28.62 $\pm$ 0.19 & 56290.17 & 15.89 $\pm$ 0.10 & 56290.19 & 14.27 $\pm$ 0.08 \\
 & 57340.50 & 12.53 $\pm$ 0.65 & 56308.06 & 28.73 $\pm$ 0.20 & 56297.11 & 16.49 $\pm$ 0.09 & 56291.08 & 14.10 $\pm$ 0.07 \\
 & 57368.53 & 11.59 $\pm$ 0.54 & 56536.21 & 28.50 $\pm$ 0.19 & 56306.10 & 16.41 $\pm$ 0.09 & 56297.13 & 14.45 $\pm$ 0.08 \\
 & 57627.76 & 11.54 $\pm$ 0.49 & 56543.23 & 28.65 $\pm$ 0.19 & 56538.26 & 16.92 $\pm$ 0.10 & 56310.05 & 13.96 $\pm$ 0.07 \\
		\hline
	\end{tabular}
\end{table*} 

\subsection{Lag Calculations} \label{sec:Lags}
We calculated the \ion{C}{iv} time lags using the interpolated cross correlation function python code PyCCF \citep{PyCCF}.  This code follows the methodology of \citet{Peterson1998} which will be summarised as follows.  The spectral line light curve is shifted by the time lag being tested and the continuum light curve is then interpolated to these adjusted time data points.  The time step between tested lags we use in this analysis is three days although tests show that the results are not sensitive to the value chosen.  We cross correlate the light curves and use the centroid of the cross-correlation function (CCF) to represent the measured time lag.  To determine the centroid time lag we find the peak correlation coefficient, $r_{\rm{max}}$, and include all values in the CCF for which $r > 0.8r_{\rm{max}}$. We chose the centroid lag measurement $\tau_{\rm{\rm{cent}}}$ to represent the time lag as opposed to the peak,  $\tau_{\rm{peak}}$, given that $\tau_{\rm{cent}}$ has been shown to better encapsulate the extent of the BLR \citep{Gaskell1986, Robinson1990}. 

In order to calculate the uncertainties on the lag measurement flux we implemented flux randomisation and random subset sampling.  Flux randomisation (FR) accounts for the uncertainties in the flux measurements by modifying the flux used in the cross correlation by drawing a new flux from a Gaussian distribution based on the flux and its standard deviation.  To assess the impact that data sampling has on the lag measurements we also use the PyCCF code's random subset sampling (RSS) which chooses a random subset of the data ($\sim$ 37 \% smaller than the original set) on which to perform the lag calculation.  In this analysis we used 10,000 realisations for the FR/RSS. We then formed a cross correlation centroid distribution (CCCD) from the lags measured from each realisation.  We measure the overall time lag from the median of the CCCD and the uncertainties are the limits such that $\tau_{\rm{cent}} \pm \Delta \tau$ contains 68.27\% of the data.  In order for a lag measurement to be added to the CCCD the peak of the CCF must be greater than 0.5.  If the peak of the CCF lies below this value the realisation is said to have failed. 

\subsection{OzDES Lag Measurements}
\begin{figure*}
	\centering
    \includegraphics[width=0.45\linewidth]{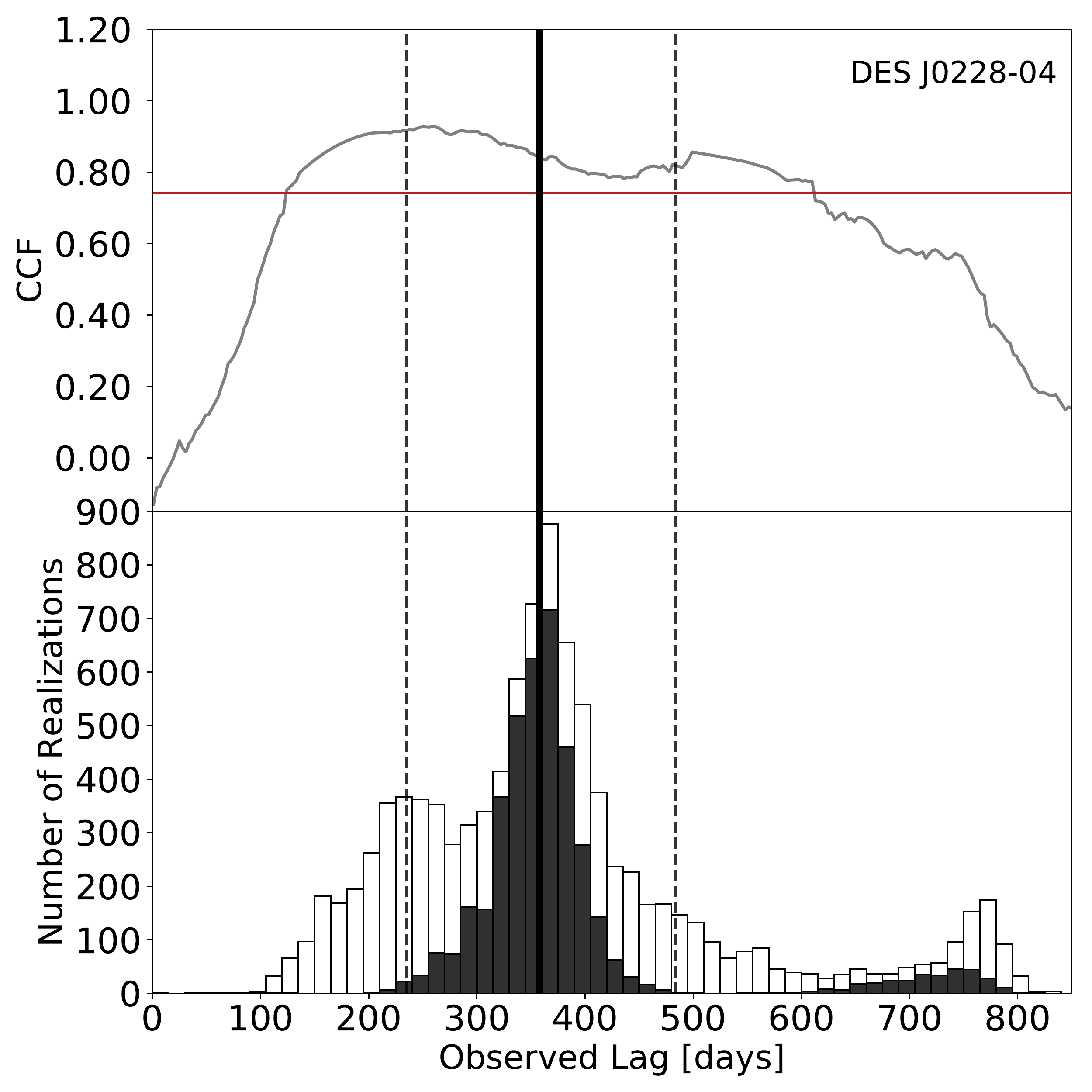}
    \includegraphics[width=0.45\linewidth]{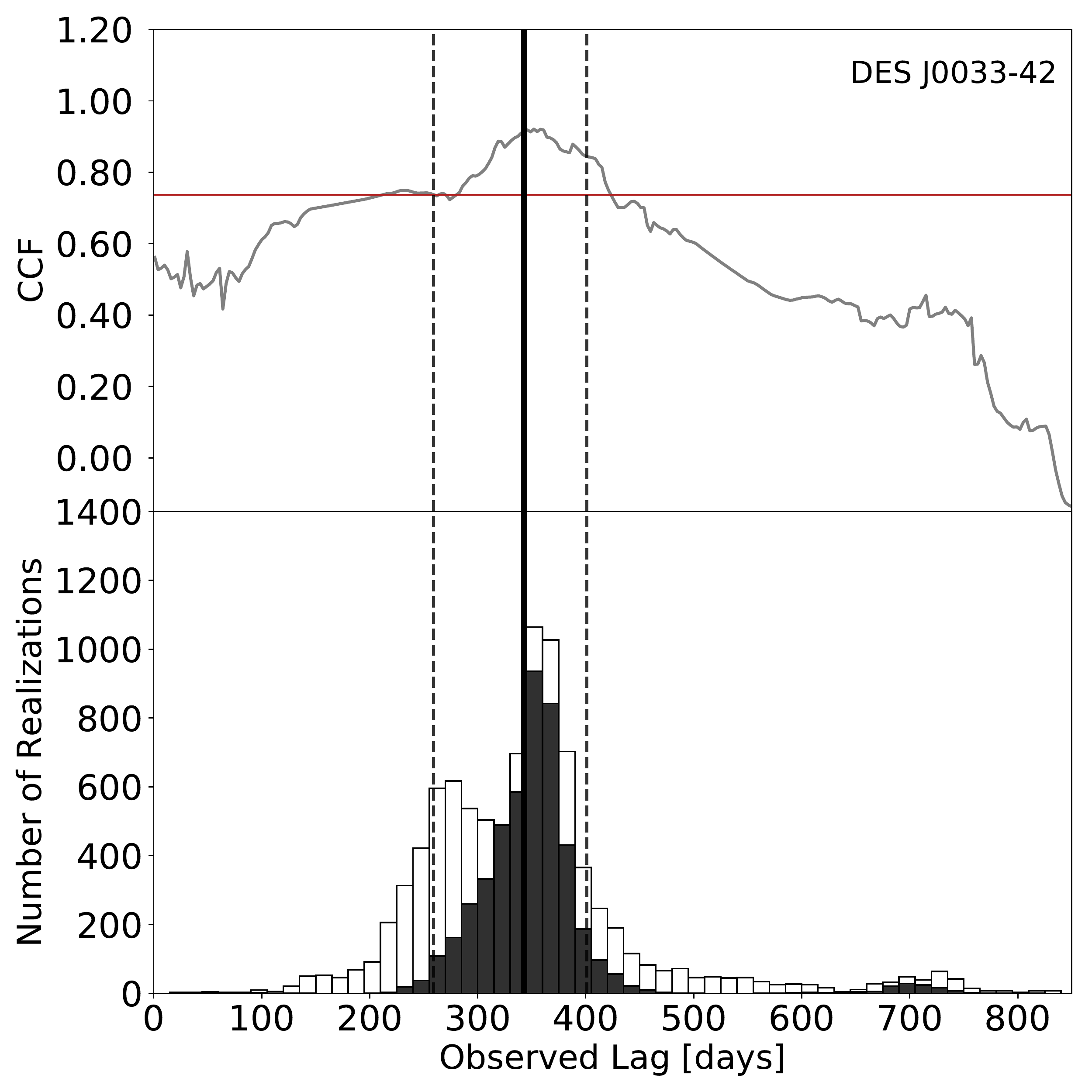}
    \caption{Lag calculations of the two OzDES AGN.   The CCF for each AGN are shown in the {\it top panel}.  The centroid of the CCF above the red line at $0.8 r_{\rm{max}}$ is used to determine the time lag for each realisation.  The final time lag is the median of the CCCD ({\it bottom panel}) which is determined using FR/RSS on the light curves with 10,000 realisations. The solid black line represents the measured time lag and the dotted lines mark the 68\% confidence levels.  The dark grey histogram is the CCCD if we down-weight lags that correspond to light curves with minimal overlap due to the seasonal gaps. The time lags we report are based on the original CCCD (white).\label{fig:iccf}}
\end{figure*}

The results of PyCCF for the two OzDES AGN are shown in Figure \ref{fig:iccf}.  The CCF is shown in the top panel and the horizontal red line shows the  $0.8r_{\rm{max}}$ value for this CCF.  The centroid of the data found above this line is used to represent the time lag,  $\tau_{\rm{cent}}$, for this realisation.  The CCCD from the $\tau_{\rm{cent}}$ values for 10,000 realisation using FR/RSS is shown in the bottom panel.  Only 0.3\% of the realisations for DES J0228-04 failed to recover a lag resulting in an observed lag of $\tau_\mathrm{\rm{obs}} = 358^{+126}_{-123}$ days.  The analysis for DES J0033-42 only failed to recover a lag for 7\% the realisations with a resulting time lag of  $\tau_\mathrm{\rm{obs}} = 343^{+58}_{-84}$ days.  The results of this analysis are summarised in Table \ref{tab:lagResults}.  We ran the analysis using the r and i bands and the results are consistent within error with those obtained using the g-band.  Previous studies have reported lag measurements with failure rates of 17\% or even higher \citep{Kaspi2007,Lira2018}. Even if we raise the correlation coefficient threshold to 0.7, less than 10\% of the realisations fail for both AGN.  Our two lag measurements are consequently of higher significance than some previous measurements. The bottom panel also shows the CCCD which has been down-weighted by the overlap between the light curves considering both survey length and the seasonal gaps.  This procedure will be described in detail in Section \ref{sec:LagPrior}.  

\begin{table*}
	\centering
	\caption{Parameters of the two AGN in this study.}
	\label{tab:lagResults}
	\begin{tabular}{|lccccccc}
		\hline
		AGN & $z$ &$\tau_{\rm{obs}}$ [days] & $\tau_{\rm{RF}}$ [days] & $\log \lambda L_{\lambda}$ [ergs s$^{-1}$] & $\sigma_{\rm{RMS}}$ [km s$^{-1}$] & M$_{\rm{BH}}$ [10$^9$ M\textsubscript{\(\odot\)}] & M$_{\rm{VP}}$ [10$^9$ M\textsubscript{\(\odot\)}] \\
		\hline
		DES J0228-04 & 1.905 & $358^{+126}_{-123}$ & $123^{+43}_{-42}$ & 46.43 $\pm$ 0.04 & 6365 $\pm$ 66 & $4.4^{+2.0}_{-1.9}$  & $1.0^{+0.3}_{-0.3}$\\[3pt]
        
		DES J0033-42 & 2.593 & $343^{+58}_{-84}$ & $95^{+16}_{-23}$ & 46.51 $\pm$ 0.02 & 6250 $\pm$ 64 & $3.3^{+1.1}_{-1.2}$& $0.7^{+0.1}_{-0.2}$ \\
		\hline
	\end{tabular}
\end{table*} 

\subsection{Method Justification}

\begin{figure*}
	\centering
    \includegraphics[width=0.45\linewidth]{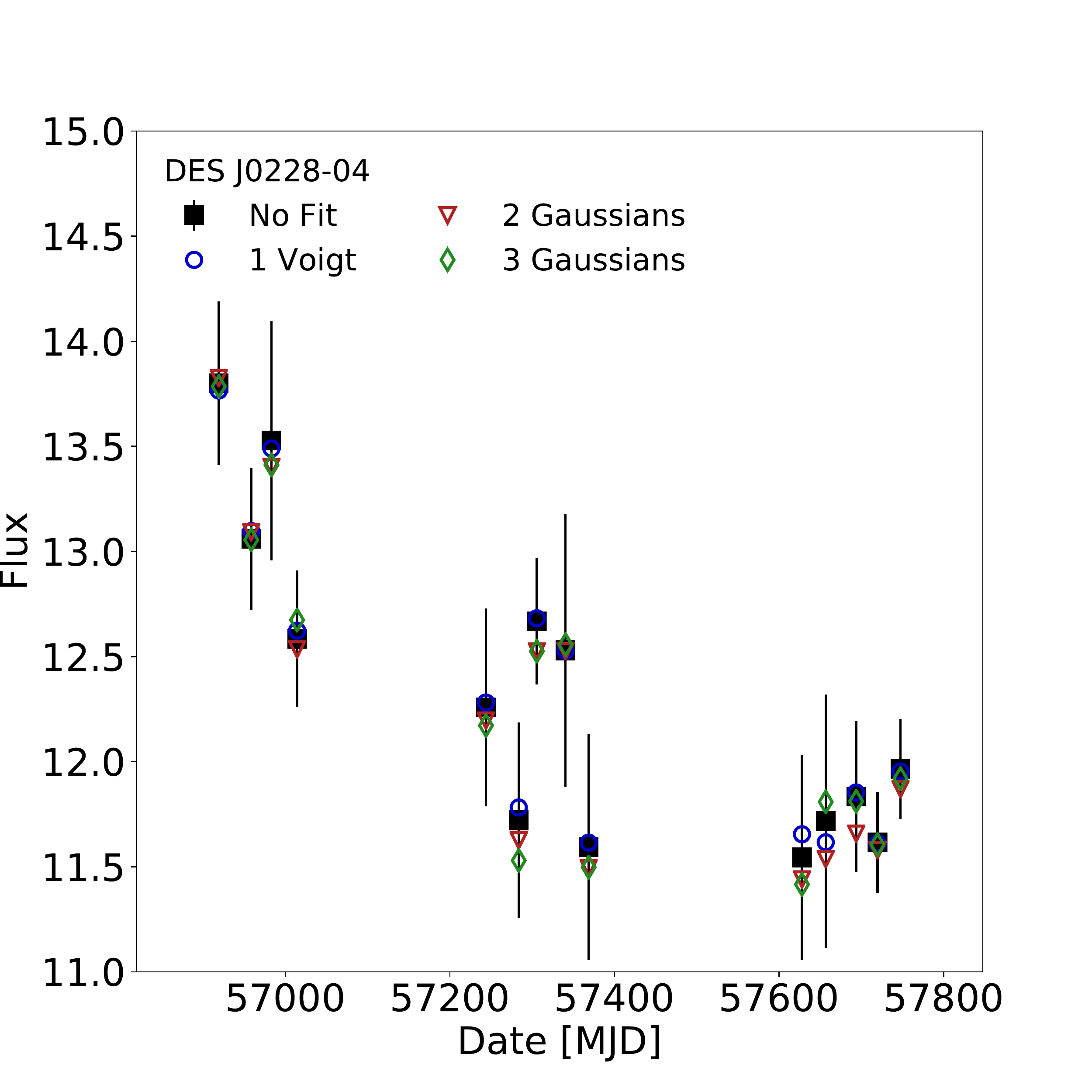}
    \includegraphics[width=0.45\linewidth]{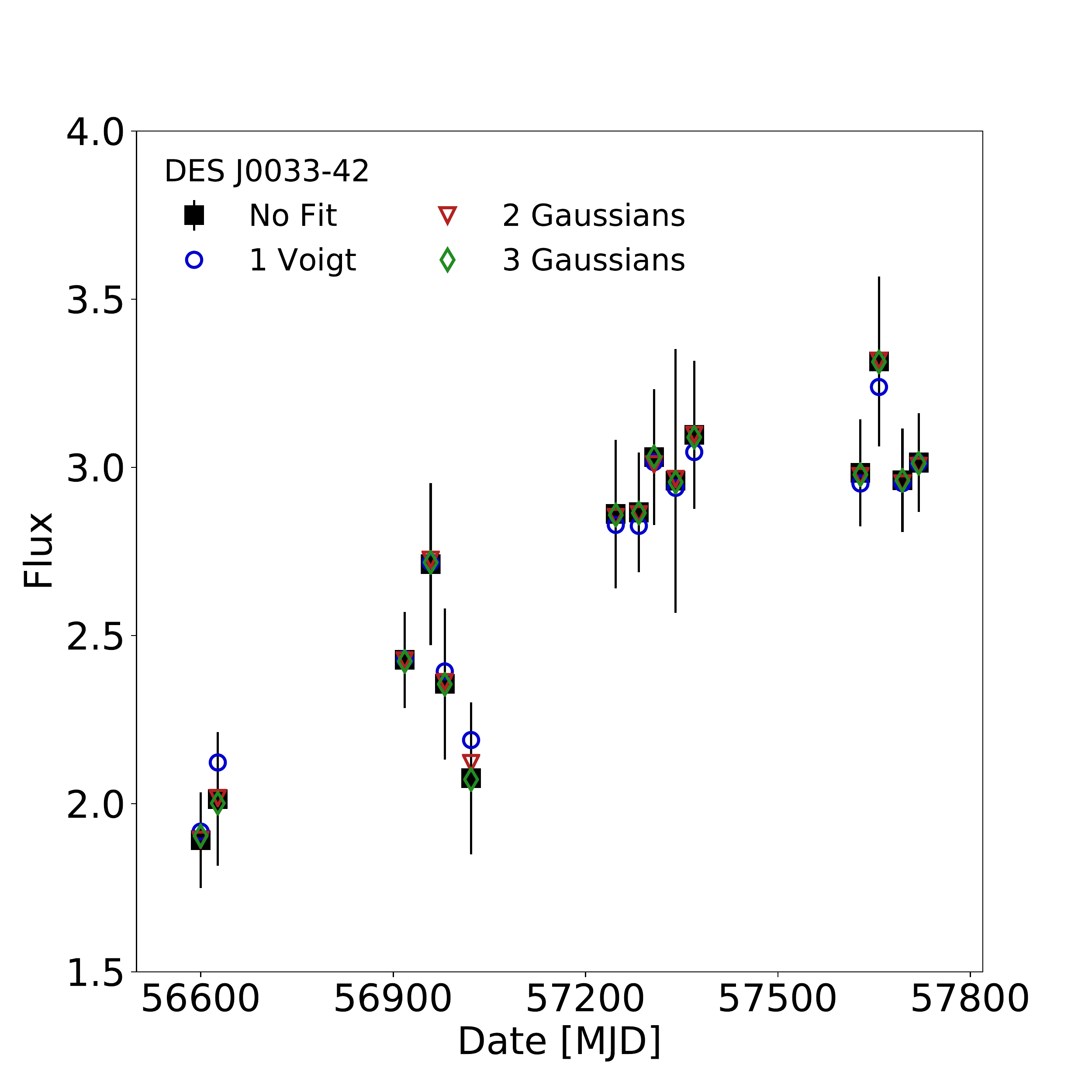}
    \caption{Results from the test to determine the validity of directly integrating the calibrated spectra.  Light curves measured from direct integration are shown by black squares with error bars. Line fluxes calculated after fitting the spectra with various line profiles lie within these uncertainties (Voigt profile - blue circle, double Gaussian - red triangle, triple Gaussian - green diamond).  Fluxes are in units of $10^{-17}$~ergs~s$^{-1}$~cm$^{-2}${\AA}$^{-1}$. \label{fig:fitTest} }
\end{figure*}

\subsubsection{Direct Integration vs Line Fitting}
We measure the line flux density by directly integrating the spectrum within the wavelength window described in Section \ref{sec:Fluxes} (red dashed lines in Figure \ref{fig:spectra}). We also experimented with parametrised fits to the emission lines. The line fitting was done after continuum subtraction and individual components of composite fits were allowed to be offset from each other in order to encapsulate any asymmetries in the wings.  The effect line fitting has on the shape of the emission line light curve is shown in Figure \ref{fig:fitTest}.  The black squares show the \ion{C}{iv} light curve for each of the AGN used in this study and their corresponding uncertainties.  The line flux measurements are also plotted after the emission line is fitted with different functions.  The functions shown here are a Voigt profile (blue circles), a double Gaussian (red triangle), and a triple Gaussian (green diamond).  The line fluxes calculated with these different fits lie within the error bars of the line flux calculated by directly integrating the emission line.  Furthermore the resulting lag measurements using these fits differ by only a few days, well within the uncertainties of the lag measurements.  This indicates that the parametrised fits to the emission line do not provide any significant benefit.  

\begin{figure*}
	\centering
    \includegraphics[width=0.45\linewidth]{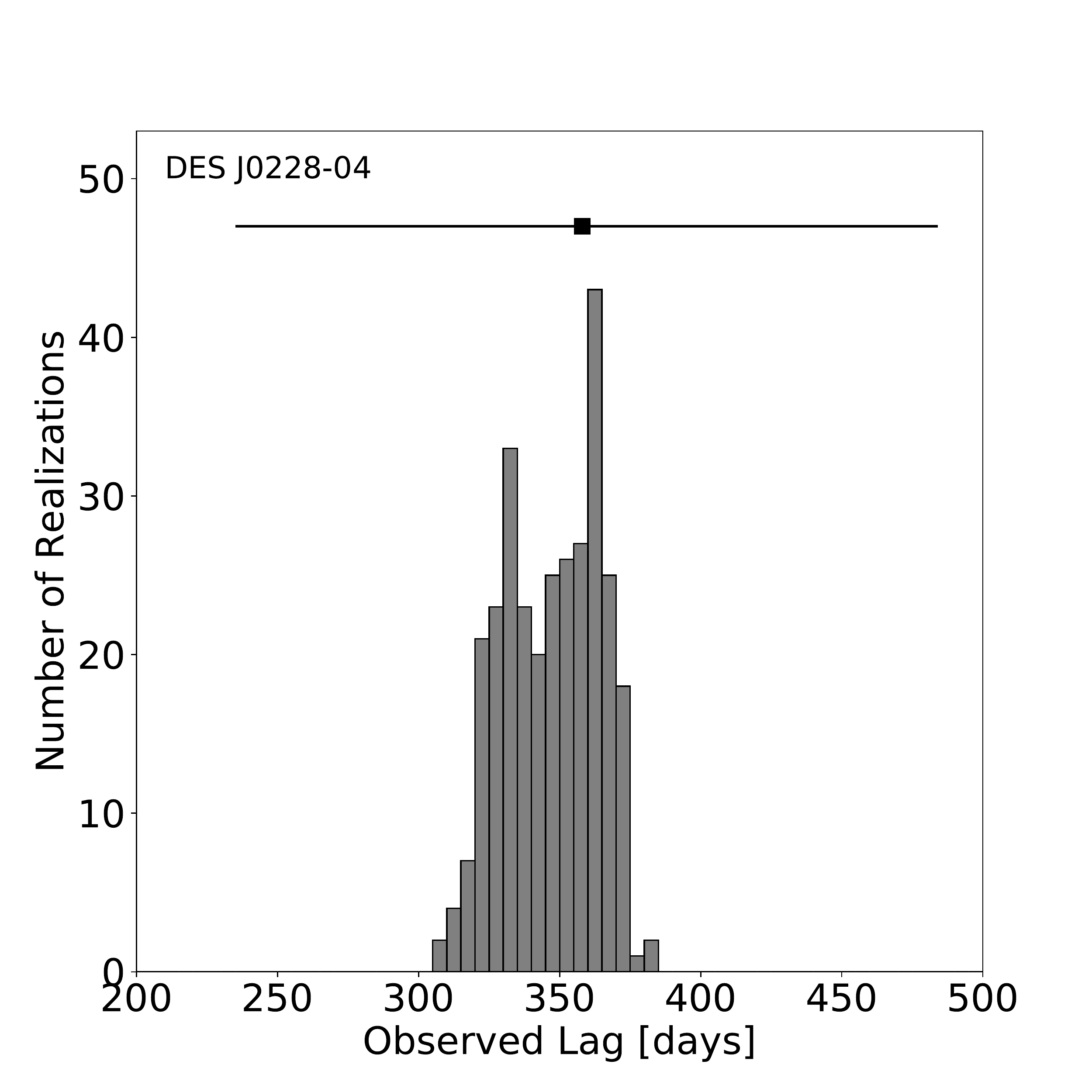}
    \includegraphics[width=0.45\linewidth]{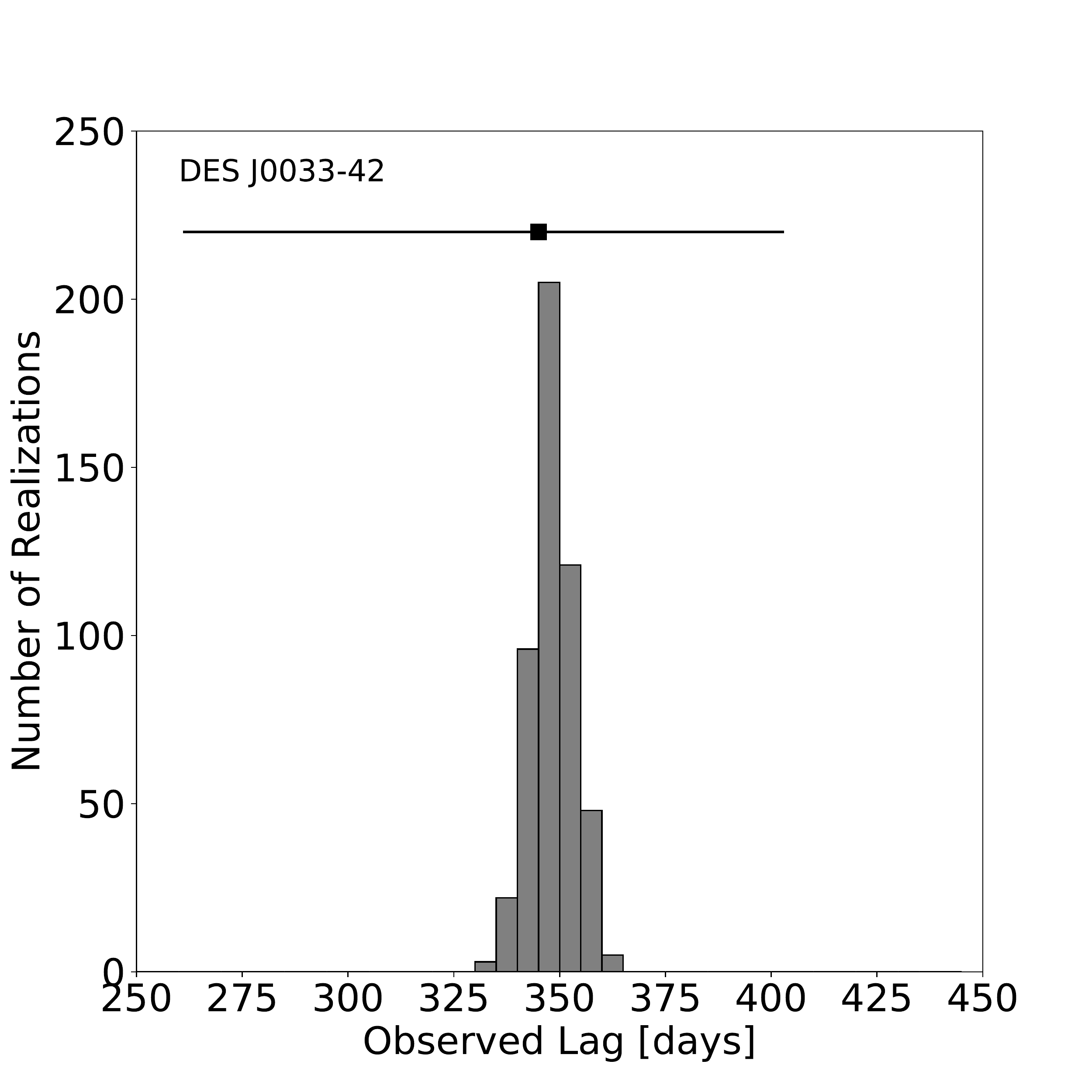}
    \caption{Results from test to determine the significance of the choice of integration window. Distributions of the lags, $\tau_{\rm{cent}}$, resulting from a sampling of 500 random integration windows.  The data point shows the lag value and corresponding uncertainty using the integration window used in this paper (1470-1595{\AA}).  The distribution of lags resulting from the various integration windows falls within the uncertainty associated with the lag measurement. \label{fig:winTest}}
\end{figure*}

\begin{figure*}
	\centering
    \includegraphics[width=0.47\linewidth]{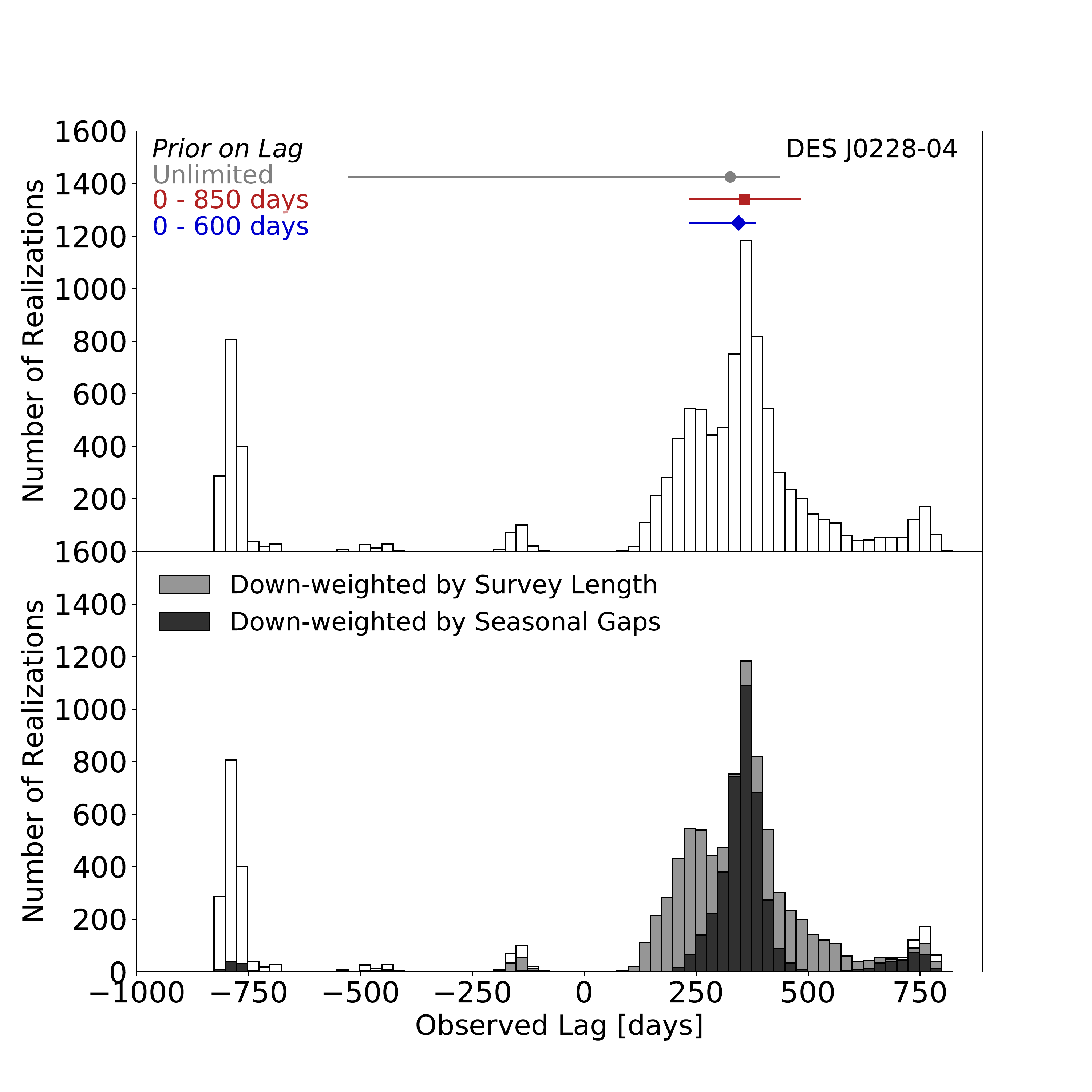}
    \includegraphics[width=0.47\linewidth]{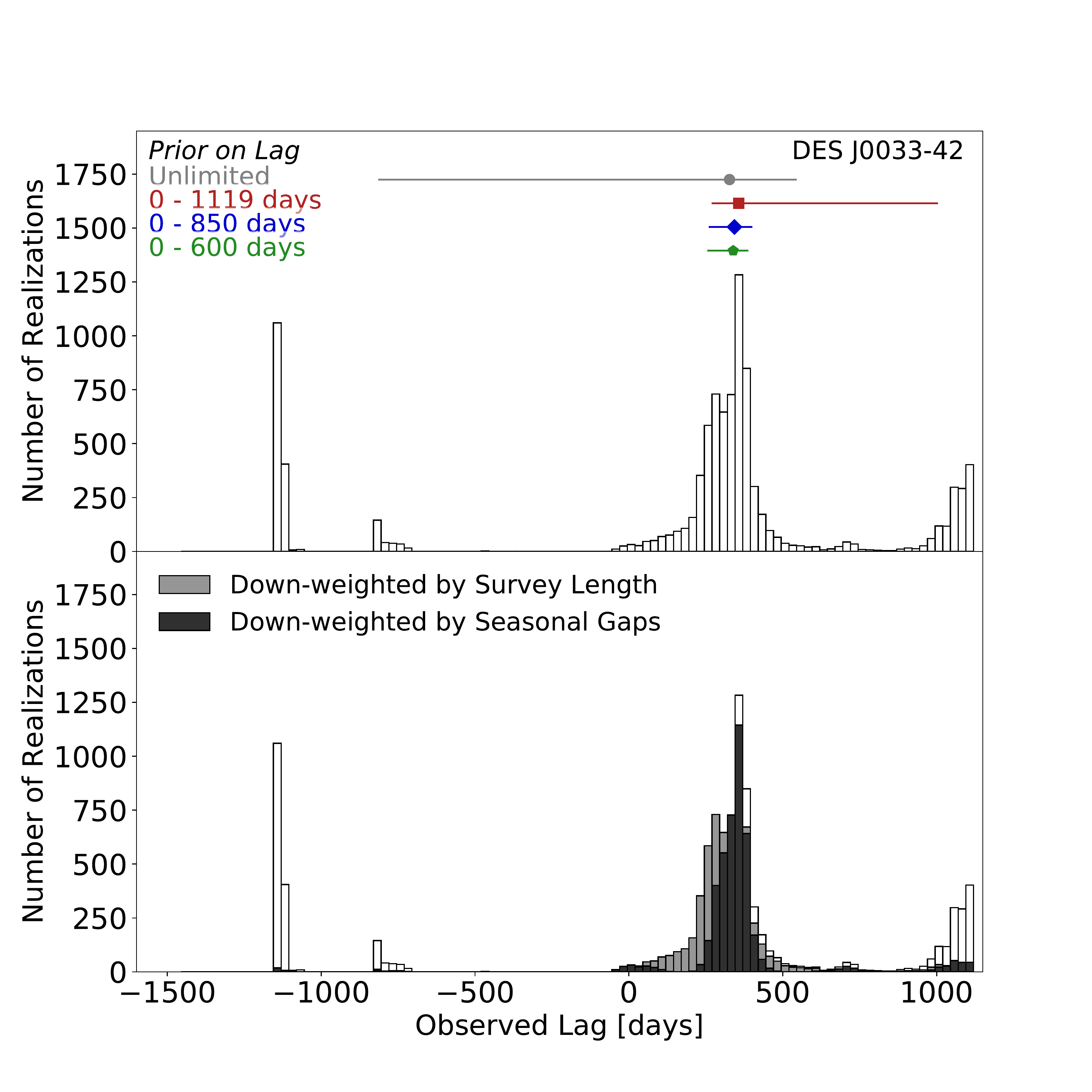}
    \caption{Results from the test to determine the effect of the time lag prior on the measured lag.  The {\it top panel} shows the effect of the prior on the lag measurement. The white histogram shows the CCCD with no prior placed on the time lag. The lag measurements and uncertainties are shown for different lag priors by the data points above the CCCD.  The {\it bottom panels} show which portions of the CCCD are affected when down-weighting for survey length (light grey) and including the seasonal gaps (dark grey).    \label{fig:rangeTest}}
\end{figure*}

\subsubsection{Line Integration Window}
It is also important to determine the significance of the choice of line integration window on the lag measurement, particularly given the presence of the red shelf so close to the \ion{C}{iv} line.  We quantified this by randomly selecting an integration window with a left bound in the range of 1456-1506{\AA} and a right bound in the range of 1570-1634{\AA}.  This tests the effects of introducing more continuum emission, including the red shelf, and also cutting off the wings of the emission line in order to avoid features not part of the \ion{C}{iv} line.   After randomly choosing 500 new integration windows we remade the light curves which we then used to measure the time lag,  $\tau_{\rm{cent}}$.   The results of this study can be seen in Figure \ref{fig:winTest}.  While there is some distribution in the recovered lag values, the result is always smaller than the uncertainties in the lag measurement.  We therefore conclude that the choice of integration window does not significantly affect our results.  As mentioned previously the OzDES spectra are made by splicing the red and blue arm of the spectrograph around $\sim$ 5700{\AA}. During Year 1 some of the spectra show a discontinuity where the splice occurs. OzDES procedures were modified from Year 2 onward to mitigate the effect. For DES J0033-42 this splice occurs at the rest frame wavelength of 1586{\AA} which is within the integration window.  However, the results shown in Figure \ref{fig:winTest} indicate that this splicing does not significantly bias our results as the randomly selected integration windows include ones which fully exclude the splice.

\begin{figure*}
	\centering
    \includegraphics[width=0.45\linewidth]{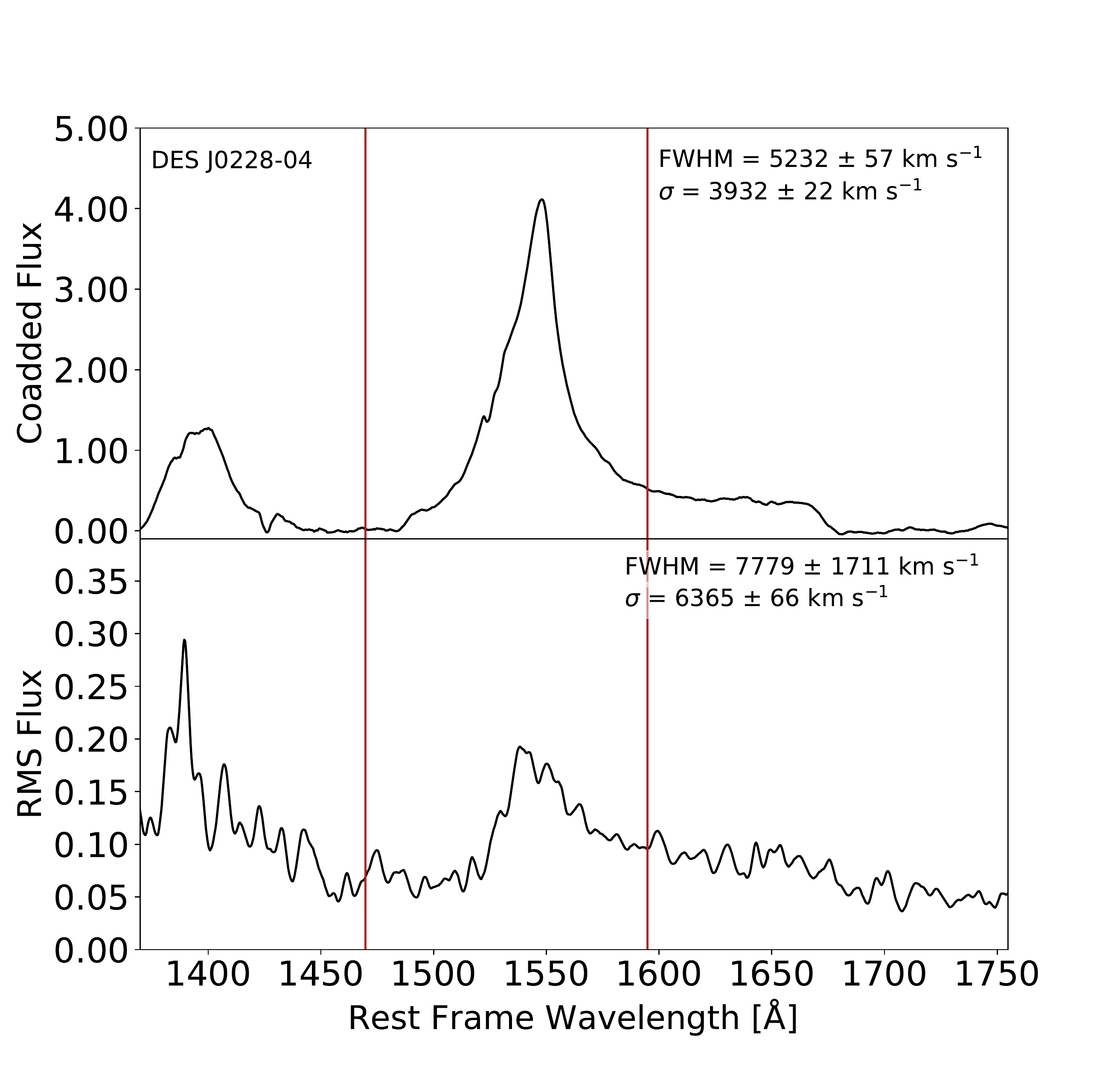}
    \includegraphics[width=0.45\linewidth]{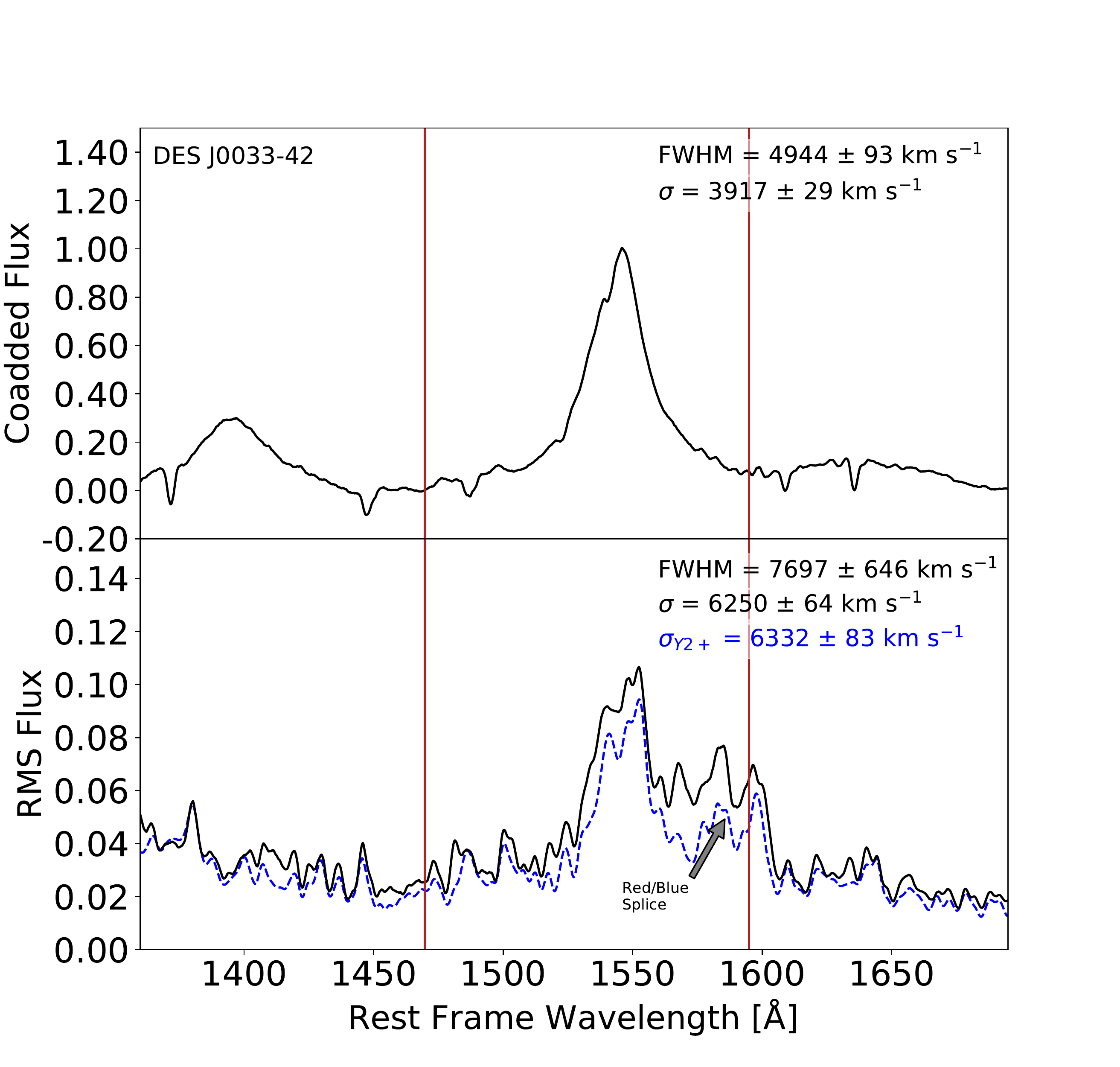}
    \caption{Smoothed coadded spectrum ({\it top panel}) and smoothed RMS spectrum ({\it bottom panel}) for each of the AGN used to calculate both the FWHM and velocity dispersion. The line integration window is indicated by the red lines. All fluxes are in units of $10^{-16}$~ergs~s$^{-1}$~cm$^{-2}${\AA}$^{-1}$. The plateau on the red side of the line for DES J0033-42 is due to the splicing of the red/blue arms of the spectrograph at 5700{\AA} in the observed frame which is more pronounced during the first two years of observation. The blue line shows the RMS spectrum with the first year removed with a corresponding velocity dispersion of $\sigma_{\rm{Y2+}} = 6332 \pm 83$ km s$^{-1}$. \label{fig:lineWidth}}
\end{figure*}

\subsubsection{Time Lag Prior} \label{sec:LagPrior}
One assumption that can have a significant effect on the measured lag as well as the uncertainties is the prior placed on the time lag.  This study assumes the time lag will fall between 0-850 days. We have tested this assumption in three ways.

\textit{Test 1:} We begin by determining the CCCD without any lag prior.  This is shown by the white histogram in the top panel of Figure \ref{fig:rangeTest}.  The corresponding time lag measurements are shown by the grey data point above the CCCD.  However, if we put a prior on the time lag, i.e. that the lag should be positive, the resulting time lag and uncertainties can be significantly affected.  For example, for both AGN the prior enforcing the assumption of a positive lag is shown in red.  The upper limit for each AGN (850 days for DES J0228-04 and 1119 days for DES J0033-42) is determined by the baseline of the observations.  If the \ion{C}{iv} line light curve is shifted by a lag value greater than this there will be no overlap between the continuum and line light curves meaning any lag measurements are primarily dependent on extrapolation, not measured data, and are therefore unreliable.  We also tested smaller lag prior ranges.

For all the lag ranges shown, over 50\% of the original CCCD falls within these values.  This indicates that a significant amount of data is not excluded by the choice of lag prior and any resulting lag measurements are reasonable \citep[see e.g.][]{Lira2018}. For DES J0228-04 over 70\% of the original CCCD falls within the smallest lag range (0-600 days) while for DES J0033-42 65\% falls within this range.

\textit{Test 2:} We performed a second test to motivate our lag prior choice using a weighting scheme derived by \citet{Grier2017} to down-weight lags that occur where there is little overlap between the two light curves after adjusting for the time lag.  Following this prescription each lag in the CCCD is weighted by the function $w = (N_{\rm{lag}}/N_0)^2$.  The number of overlapping data points with no lag correction is given by $N_0$ and the number of overlapping data points after lag correction is $N_{\rm{lag}}$.   Lags with a low number of overlapping data points rely heavily on the extrapolation assumed, not the real measured data points.  This makes these lag values less reliable.  The bottom panel of Figure \ref{fig:rangeTest} shows the original CCCD in white with the CCCD down-weighted using this prescription to account for the length of the survey shown by the light grey region.  This eliminates much of the data at negative lags and the slight bump after 1000 days seen for DES J0033-42.  

\textit{Test 3:} This original weighting scheme only takes into account the baseline of the OzDES observations as this is most directly related to the prior we can place on the lag measurements.  However, an important feature of the OzDES survey is the seasonal gaps of around 200 days between each observing season.   In order to test the significance the seasonal gaps have on our ability to recover time lags we modified the weighting procedure to account for the case when there is no overlap between the light curves because the lag moves the line light curve into the seasonal gaps of the continuum light curves. This is shown by the dark grey histogram which significantly decreases the width of the main peak.  

As a result of this analysis we choose to report our results using the lag prior of 0-850 days.  This is shown by the white histogram on the lower panel of Figure \ref{fig:iccf}.  On Figure \ref{fig:iccf} we also plot the CCCD after down-weighting lags based on survey length and seasonal gaps (dark grey histogram).  We use the unweighted CCCD for our final result in order to avoid biasing the analysis by only considering lags to which the survey is sensitive.  However, the down-weighted CCCD does shed light into the importance of the cadence of the observations.  Specifically, the peaks in the CCCD around 200 days are likely due to the seasonal gaps in OzDES which shows that survey cadence is an important limiting factor and source of uncertainty in lag recovery.

\begin{figure*}
	\includegraphics[width=.7\linewidth]{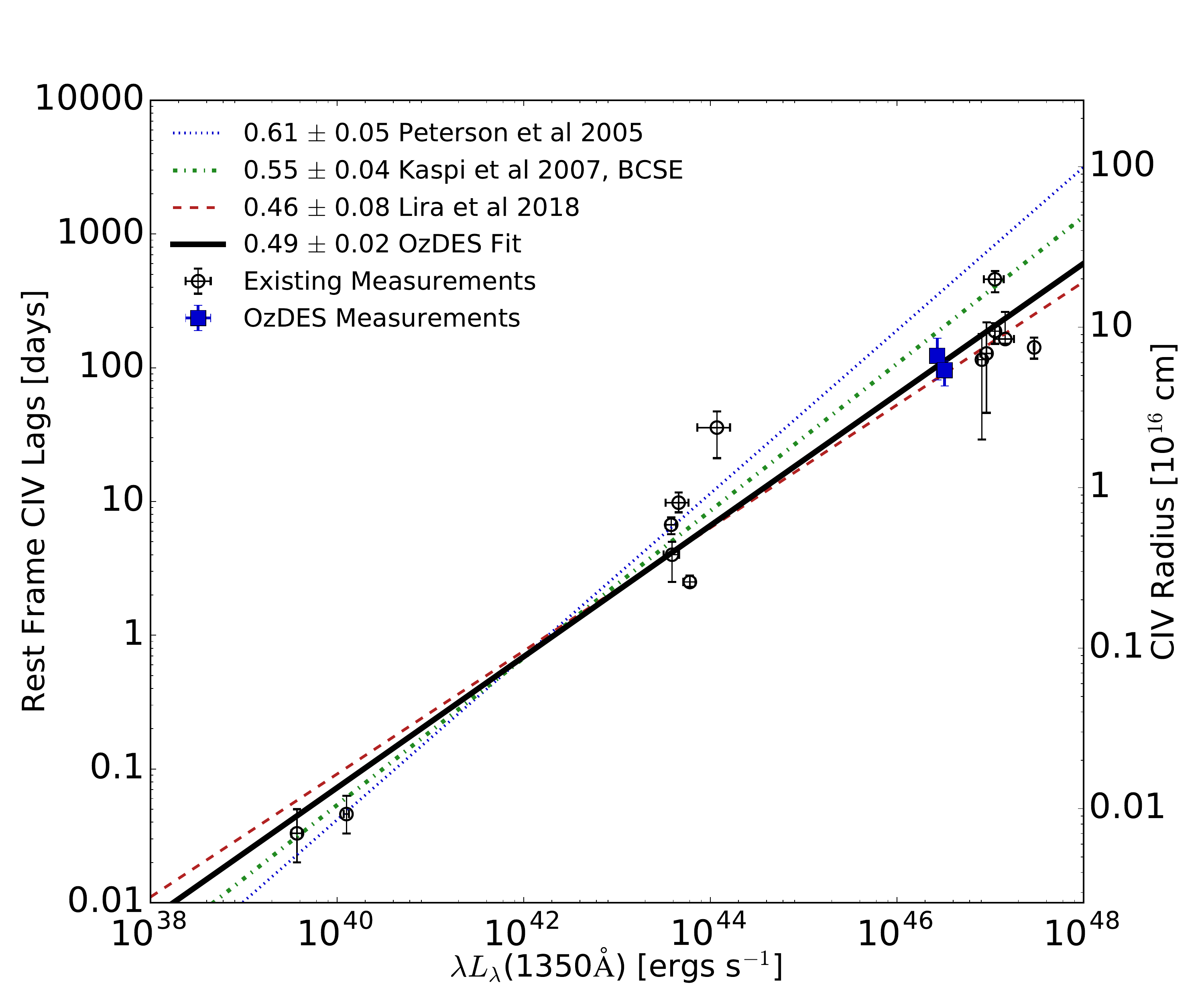}
    \caption{Rest frame \ion{C}{iv} lag versus the AGN luminosity $\lambda L_{\lambda}$(1350{\AA}).  Previous lag measurements \citep{Peterson2005,Kaspi2007,Trevese2014,Lira2018} are indicated with black circles and the OzDES measurements are shown in blue squares.  Table \ref{tab:existingLags} summarises the data used in this figure. The best fit line to the entire data set is the black line with a slope of $0.49 \pm 0.02$ found using the BCES method.  Best fit lines obtained by \citet{Peterson2005} (blue dotted line) including the low luminosity sources and the \citet{Kaspi2007} BCES result (green dashed dotted) which extended the fit to higher luminosities are shown for comparison. The red dashed line corresponds to that measured by \citet{Lira2018}.  The corresponding \ion{C}{iv} BLR radius in cm is also shown.\label{fig:RLDiagram}}
\end{figure*}   

\section{Results} \label{sec:Results}
\subsection{Black Hole Mass}
We need to know the velocity distribution of the BLR clouds in order to determine the black hole mass.  The velocity can be measured from either the mean spectrum or the root mean squared (RMS) spectrum, which isolates the reverberating components of the emission line.   A smoothed version of both the mean and RMS spectrum for each AGN is shown in Figure \ref{fig:lineWidth}. The velocity can be characterised by either the velocity dispersion ($\sigma$) or the full width half maximum (FWHM).  Figure \ref{fig:lineWidth} includes the FWHM and $\sigma$ values for each spectrum calculated within the line integration window indicated by the vertical red lines.  The line widths have been corrected for the resolution of the spectrograph which is $\sim 3~\text{\AA}$.  For our black hole mass calculations we used the line dispersion, $\sigma$, of the RMS spectrum to measure the velocity of the BLR clouds, as work by \citet{Peterson2004b} concluded that the line dispersion is the best representation of the BLR velocity. We calculated the uncertainties by performing a bootstrapping procedure similar to that described above to account for the impact of continuum subtraction and uncertainties on the flux values before calculating the line width.  Note that the RMS spectrum for DES J0033-42 (bottom right panel) has a distinct plateau on the red side of the line.  This is due to the splicing between the red and blue arms of the spectrograph at $\sim$ 5700 {\AA}.  The plateau is less pronounced if the spectra from Year 1 are not included in the calculation of the RMS spectrum.  Regardless of whether or not the Year 1 data is included the measurement of the velocity dispersion does not change significantly. This is shown by the blue line in the right panel of Figure \ref{fig:lineWidth}.  There is still some increase in flux around the splice.  If the right edge of the integration window within which the line width is calculated is shifted to completely exclude the splice (1580 {\AA} in the rest frame) the line width decreases by around 650 km s$^{-1}$ for both AGN and the resulting black hole mass is consistent with those obtained when considering the full integration window. This indicates that the effect of the red/blue arm splice is a subdominant effect here but a more rigorous analysis of the line width measurements will be done in the future.

The black hole masses are $4.4^{+2.0}_{-1.9}$$\times$10$^9~M\textsubscript{\(\odot\)}$ for DES J0228-04 and $3.3^{+1.1}_{-1.2}$$\times$10$^9~M\textsubscript{\(\odot\)}$ for DES J0033-42 (see Table~\ref{tab:lagResults}).  We calculated the masses using Equation \ref{eq:bhmass} and $f=4.47~\pm~1.25$ \citep{Woo2015}.

\subsection{Radius-Luminosity Relationship}
Previous radius-luminosity relationships using the \ion{C}{iv} line measurements  were made using the $\lambda L_{\lambda}(1350~\text{\AA})$ luminosity \citep{Peterson2005,Kaspi2007,Trevese2014,Lira2018}. They were generally found using the Bivariate Correlated Errors and Intrinsic Scatter (BCES) method developed by \citet{Akritas1996} to account for uncertainties in both luminosity and radius. For low luminosity AGN, \citet{Peterson2005} found a slope of $0.61 \pm 0.05$.  Including higher luminosity measurements \citet{Kaspi2007} found a slope of $0.55 \pm 0.04$ and $0.52 \pm 0.04$ using the FITEXY method to iteratively minimise  $\chi^2$ \citep{Press1992} combined with the intrinsic scatter prescription of \citet{Tremaine2002}. Most recently \citet{Lira2018} published a $R-L$ slope of $0.46 \pm 0.08$. We have constructed a new $R-L$ relationship by adding our two OzDES lag measurements to those in \citet{Kaspi2007}, \citet{Trevese2014}, and the four most robust \ion{C}{iv} lag measurements (CT286, CT406, J214355, J221516) made by \citet{Lira2018}.  Those four were measured at a 1$\sigma$ level with a $<60\%$ failure rate.  The results we used from the literature are summarised in Table \ref{tab:existingLags}. We calculated the $\lambda L_{\lambda}(1350~\text{\AA})$ luminosity for the OzDES AGN assuming $H_0$~=~70~km~s$^{-1}$~Mpc$^{-1}$, $\Omega_M$~=~0.3, and $\Omega_\Lambda$~=~0.7. The resulting $R-L$ diagram can be seen in Figure \ref{fig:RLDiagram}.  We find that the best-fit slope is $0.49 \pm 0.02$ for the following R-L relationship:

\begin{equation}
\log R (\mathrm{lt-days}) = (0.82 \pm 0.08) + (0.49 \pm 0.02)\log\left(\frac{\lambda L_{\lambda}(1350~\text{\AA})}{10^{44}\ \text{ergs s}^{-1}} \right),
\end{equation}
based on the public BCES code by \citet{Nemmen2012}. As the BCES method does not account for asymmetry in the error bars we calculated the slope using both the upper and lower values for uncertainties in the radius.  This did not affect the resulting slope, intercept, or corresponding uncertainties.  Our results are also consistent with a Monte Carlo linear regression procedure which varies the data points within their asymmetrical error bars.  While yielding the same slope the $R-L$ relationship using the existing data summarised in Table \ref{tab:lagResults} but without including the two OzDES AGN has a larger error of $0.03$.  

When comparing our measured time lags with those estimated from previous $R-L$ relationships the lags measured by \citet{Kaspi2007} are 50\% and 110\% larger for DES J0228-04 and DES J0033-42 respectively.  The slope derived in \citet{Kaspi2007} is consistent with the high redshift relationship $R(H \beta ) \propto \lambda L_{\lambda}(1350~\text{\AA}) ^{0.56}$ \citep{Kaspi2005}.  However, this slope was using a data set which only included one high luminosity measurement. Our lags are consistent within the measured error bars with the lags predicted by the most recent $R-L$ relationship found by \citet{Lira2018} which included the higher luminosity AGN.  Using the dispersion based single epoch mass estimate for \ion{C}{iv} found by \citet{Vestergaard2006} the black hole masses are $4.2$$\times$10$^9~M\textsubscript{\(\odot\)}$ for DES J0228-04 and $4.5$$\times$10$^9~M\textsubscript{\(\odot\)}$ for DES J0033-42 which are comparable with the masses derived in this paper.

\begin{table}
	\centering
	\caption{Rest frame time lags and 1350 \text{\AA} luminosities for all \ion{C}{iv} lags used to derive the $R-L$ relationship.}
	\label{tab:existingLags}
	\begin{tabular}{|lcccccc}
		\hline
		AGN & $\log \lambda L_{\lambda}$ [ergs s$^{-1}$] &  $\tau_{\rm{RF}}$ [days] & Ref. \\
		\hline
		DES J0228-04  & 46.43 $\pm$ 0.04  & $123^{+43}_{-42}$ &  1 \\[3pt]
        
		DES J0033-42 & 46.51 $\pm$ 0.02  & $95^{+16}_{-23}$ &  1 \\[3pt]
        
		NGC 4395 Visit 2 & 39.57 $\pm$ 0.06  & $0.033^{+0.017}_{-0.013}$ &  2 \\[3pt]
        
		NGC 4395 Visit 3 & 40.10 $\pm$ 0.03  & $0.046^{+0.017}_{-0.013}$ &  2  \\[3pt]
        
		NGC 3783 & 43.59 $\pm$ 0.09  & $4.0^{+1.0}_{-1.5}$ &  2 \\[3pt]
        
		NGC 5548 Year 1 & 43.66 $\pm$ 0.14  & $9.8^{+1.9}_{-1.5}$ &  2 \\[3pt]
        
		NGC 5548 Year 5 & 43.58 $\pm$ 0.06  & $6.7^{+0.9}_{-1.0}$ &  2 \\[3pt]
        
		NGC 7469 & 43.78 $\pm$ 0.07  & $2.5^{+0.3}_{-0.2}$ &  2 \\[3pt]
        
		3C 390.3 & 44.07 $\pm$ 0.21  & $35.7^{+11.4}_{-14.6}$ &  2 \\[3pt]
        
		S5 0836+71 & 47.05 $\pm$ 0.06  & $188^{+27}_{-37}$ &  3 \\[3pt]
        
		PG 1247+267 & 47.47 $\pm$ 0.003  & $142^{+26}_{-25}$ &  4\\[3pt]
        
		CT286 & 47.05 $\pm$ 0.12  & $459^{+71}_{-92}$ & 5 \\[3pt]
        
		CT406 & 46.91 $\pm$ 0.05  & $115^{+64}_{-86}$ &  5 \\[3pt]
        
		J214355 & 46.96 $\pm$ 0.07  & $128^{+91}_{-82}$ &  5\\[3pt]
        
		J221516 & 47.16 $\pm$ 0.12  & $165^{+98}_{-13}$ &  5\\
		\hline
	\end{tabular}
	\\
	\flushleft
    \textbf{References:} (1) This work; (2) \citet{Peterson2005} and references therein; (3) \citet{Kaspi2007}; (4) \citet{Trevese2014}; (5) \citet{Lira2018}
\end{table} 

\section{Conclusion} \label{sec:Discussion}

In this paper we present the first OzDES reverberation mapping black hole mass measurements using the \ion{C}{iv} line for some of the highest quality AGN observed during the first four years of OzDES operations.  The techniques tested and implemented here will now be used on the wider OzDES sample and will include an additional two year baseline once the data from Year 5 and 6 have been processed.  The data obtained from OzDES is complementary to its northern hemisphere counterpart, the Sloan Digital Sky Survey Reverberation Mapping program (SDSS-RM) \citep{Shen2015}.  Similar to OzDES, SDSS-RM observed 849 AGN out to $z < 4.5$.  The primary difference between them is that SDSS-RM has published results based on approximately 30 epochs of observations over 6 months where as OzDES has around 23 epochs over 6 years.

We measured black hole masses of $4.4 \times 10^{9}$ M$_\odot$ for DES J0228-04 and $3.3 \times 10^{9}$ M$_\odot$ for DES J0033-42. These are amongst the highest redshift AGN with the highest mass black holes measured to date with this technique. We have used these new measurements to update  the \ion{C}{iv} $R-L$ relationship and derive the slope of $\alpha = 0.49 \pm 0.02$.  Of the 771 AGN regularly observed with OzDES, those with \ion{C}{iv} lines range in luminosity from  $10^{44.3}\text{ ergs s}^{-1}~\leq~\lambda L_\lambda(1350~\text{\AA})~\leq~10^{47.2}\text{ ergs s}^{-1}$.  Based on this range in luminosities we expect to be able to fill in the gap between moderate luminosity AGN ($\sim$ 10$^{43}$~ergs~s$^{-1}$) and high luminosity AGN (>~10$^{46}$~ergs~s$^{-1}$) that is seen in Figure~\ref{fig:RLDiagram}.  This will provide an opportunity to investigate the source of the scatter seen in this relationship. 

In order to study the stratification of the BLR we chose the OzDES AGN candidates to include AGN where multiple emission lines that are in the observed spectroscopic bandpass.  Our data set includes 148 AGN, such as DES J0228-04, which contain both \ion{C}{iv} and \ion{Mg}{ii} and 27 AGN which include \ion{Mg}{ii} and H$\beta$.  Given the high redshift range of the OzDES RM candidates, $0.1 < z < 4.5$, we will have the opportunity to study the growth of supermassive black holes and test the theory that AGN can be used as high redshift standard candles. 

\section*{Acknowledgements}
This research was funded partially by the Australian Government through the Australian Research Council through project DP160100930.
PM and ZY were supported in part by the United States National Science Foundation under Grant No. 161553.

Based on data acquired at the Anglo-Australian Telescope, under program A/2013B/012. We acknowledge the traditional owners of the land on which the AAT stands, the Gamilaraay people, and pay our respects to elders past and present.

Funding for the DES Projects has been provided by the U.S. Department of Energy, the U.S. National Science Foundation, the Ministry of Science and Education of Spain, 
the Science and Technology Facilities Council of the United Kingdom, the Higher Education Funding Council for England, the National Center for Supercomputing 
Applications at the University of Illinois at Urbana-Champaign, the Kavli Institute of Cosmological Physics at the University of Chicago, 
the Center for Cosmology and Astro-Particle Physics at the Ohio State University,
the Mitchell Institute for Fundamental Physics and Astronomy at Texas A\&M University, Financiadora de Estudos e Projetos, 
Funda{\c c}{\~a}o Carlos Chagas Filho de Amparo {\`a} Pesquisa do Estado do Rio de Janeiro, Conselho Nacional de Desenvolvimento Cient{\'i}fico e Tecnol{\'o}gico and 
the Minist{\'e}rio da Ci{\^e}ncia, Tecnologia e Inova{\c c}{\~a}o, the Deutsche Forschungsgemeinschaft and the Collaborating Institutions in the Dark Energy Survey. 

The Collaborating Institutions are Argonne National Laboratory, the University of California at Santa Cruz, the University of Cambridge, Centro de Investigaciones Energ{\'e}ticas, 
Medioambientales y Tecnol{\'o}gicas-Madrid, the University of Chicago, University College London, the DES-Brazil Consortium, the University of Edinburgh, 
the Eidgen{\"o}ssische Technische Hochschule (ETH) Z{\"u}rich, 
Fermi National Accelerator Laboratory, the University of Illinois at Urbana-Champaign, the Institut de Ci{\`e}ncies de l'Espai (IEEC/CSIC), 
the Institut de F{\'i}sica d'Altes Energies, Lawrence Berkeley National Laboratory, the Ludwig-Maximilians Universit{\"a}t M{\"u}nchen and the associated Excellence Cluster Universe, 
the University of Michigan, the National Optical Astronomy Observatory, the University of Nottingham, The Ohio State University, the University of Pennsylvania, the University of Portsmouth, 
SLAC National Accelerator Laboratory, Stanford University, the University of Sussex, Texas A\&M University, and the OzDES Membership Consortium.

Based in part on observations at Cerro Tololo Inter-American Observatory, National Optical Astronomy Observatory, which is operated by the Association of 
Universities for Research in Astronomy (AURA) under a cooperative agreement with the National Science Foundation.

The DES data management system is supported by the National Science Foundation under Grant Numbers AST-1138766 and AST-1536171.
The DES participants from Spanish institutions are partially supported by MINECO under grants AYA2015-71825, ESP2015-66861, FPA2015-68048, SEV-2016-0588, SEV-2016-0597, and MDM-2015-0509, 
some of which include ERDF funds from the European Union. IFAE is partially funded by the CERCA program of the Generalitat de Catalunya.
Research leading to these results has received funding from the European Research
Council under the European Union's Seventh Framework Program (FP7/2007-2013) including ERC grant agreements 240672, 291329, and 306478.
We  acknowledge support from the Australian Research Council Centre of Excellence for All-sky Astrophysics (CAASTRO), through project number CE110001020, and the Brazilian Instituto Nacional de Ci\^encia
e Tecnologia (INCT) e-Universe (CNPq grant 465376/2014-2).

This manuscript has been authored by Fermi Research Alliance, LLC under Contract No. DE-AC02-07CH11359 with the U.S. Department of Energy, Office of Science, Office of High Energy Physics. The United States Government retains and the publisher, by accepting the article for publication, acknowledges that the United States Government retains a non-exclusive, paid-up, irrevocable, world-wide license to publish or reproduce the published form of this manuscript, or allow others to do so, for United States Government purposes.




\vspace{12pt}
$^{1}$ School of Mathematics and Physics, University of Queensland,  Brisbane, QLD 4072, Australia\\
$^{2}$ Center for Cosmology and Astro-Particle Physics, The Ohio State University, Columbus, OH 43210, USA\\
$^{3}$ Department of Astronomy, The Ohio State University, Columbus, OH 43210, USA\\
$^{4}$ School of Physics, University of Melbourne, Parkville, VIC 3010, Australia\\
$^{5}$ The Research School of Astronomy and Astrophysics, Australian National University, ACT 2601, Australia\\
$^{6}$ Department of Physics and Astronomy, University of California, Irvine, Irvine, CA 92697, USA\\
$^{7}$ Fermi National Accelerator Laboratory, P. O. Box 500, Batavia, IL 60510, USA\\
$^{8}$ Korea Astronomy and Space Science Institute, Yuseong-gu, Daejeon, 305-348, Korea\\
$^{9}$ Institute of Cosmology and Gravitation, University of Portsmouth, Portsmouth, PO1 3FX, UK\\
$^{10}$ Institute of Astronomy, University of Cambridge, Madingley Road, Cambridge CB3 0HA, UK\\
$^{11}$ Kavli Institute for Cosmology, University of Cambridge, Madingley Road, Cambridge CB3 0HA, UK\\
$^{12}$ Department of Physics \& Astronomy, University College London, Gower Street, London, WC1E 6BT, UK\\
$^{13}$ Kavli Institute for Particle Astrophysics \& Cosmology, P. O. Box 2450, Stanford University, Stanford, CA 94305, USA\\
$^{14}$ SLAC National Accelerator Laboratory, Menlo Park, CA 94025, USA\\
$^{15}$ Centro de Investigaciones Energ\'eticas, Medioambientales y Tecnol\'ogicas (CIEMAT), Madrid, Spain\\
$^{16}$ Laborat\'orio Interinstitucional de e-Astronomia - LIneA, Rua Gal. Jos\'e Cristino 77, Rio de Janeiro, RJ - 20921-400, Brazil\\
$^{17}$ INAF, Astrophysical Observatory of Turin, I-10025 Pino Torinese, Italy\\
$^{18}$ Department of Astronomy, University of Illinois at Urbana-Champaign, 1002 W. Green Street, Urbana, IL 61801, USA\\
$^{19}$ National Center for Supercomputing Applications, 1205 West Clark St., Urbana, IL 61801, USA\\
$^{20}$ Institut de F\'{\i}sica d'Altes Energies (IFAE), The Barcelona Institute of Science and Technology, Campus UAB, 08193 Bellaterra (Barcelona) Spain\\
$^{21}$ Institut d'Estudis Espacials de Catalunya (IEEC), 08034 Barcelona, Spain\\
$^{22}$ Institute of Space Sciences (ICE, CSIC),  Campus UAB, Carrer de Can Magrans, s/n,  08193 Barcelona, Spain\\
$^{23}$ School of Physics and Astronomy, University of Southampton,  Southampton, SO17 1BJ, UK\\
$^{24}$ Department of Physics, IIT Hyderabad, Kandi, Telangana 502285, India\\
$^{25}$ Kavli Institute for Cosmological Physics, University of Chicago, Chicago, IL 60637, USA\\
$^{26}$ Instituto de Fisica Teorica UAM/CSIC, Universidad Autonoma de Madrid, 28049 Madrid, Spain\\
$^{27}$ Department of Astronomy, University of Michigan, Ann Arbor, MI 48109, USA\\
$^{28}$ Department of Physics, University of Michigan, Ann Arbor, MI 48109, USA\\
$^{29}$ Department of Physics, Stanford University, 382 Via Pueblo Mall, Stanford, CA 94305, USA\\
$^{30}$ Department of Physics, ETH Zurich, Wolfgang-Pauli-Strasse 16, CH-8093 Zurich, Switzerland\\
$^{31}$ Santa Cruz Institute for Particle Physics, Santa Cruz, CA 95064, USA\\
$^{32}$ Department of Physics, The Ohio State University, Columbus, OH 43210, USA\\
$^{33}$ Max Planck Institute for Extraterrestrial Physics, Giessenbachstrasse, 85748 Garching, Germany\\
$^{34}$ Universit\"ats-Sternwarte, Fakult\"at f\"ur Physik, Ludwig-Maximilians Universit\"at M\"unchen, Scheinerstr. 1, 81679 M\"unchen, Germany\\
$^{35}$ Harvard-Smithsonian Center for Astrophysics, Cambridge, MA 02138, USA\\
$^{36}$ Department of Astronomy/Steward Observatory, 933 North Cherry Avenue, Tucson, AZ 85721-0065, USA\\
$^{37}$ Australian Astronomical Optics, Macquarie University, North Ryde, NSW 2113, Australia\\
$^{38}$ Sydney Institute for Astronomy, School of Physics, A28, The University of Sydney, NSW 2006, Australia\\
$^{39}$ Departamento de F\'isica Matem\'atica, Instituto de F\'isica, Universidade de S\~ao Paulo, CP 66318, S\~ao Paulo, SP, 05314-970, Brazil\\
$^{40}$ Observat\'orio Nacional, Rua Gal. Jos\'e Cristino 77, Rio de Janeiro, RJ - 20921-400, Brazil\\
$^{41}$ Instituci\'o Catalana de Recerca i Estudis Avan\c{c}ats, E-08010 Barcelona, Spain\\
$^{42}$ Department of Astrophysical Sciences, Princeton University, Peyton Hall, Princeton, NJ 08544, USA\\
$^{43}$ Department of Physics and Astronomy, Pevensey Building, University of Sussex, Brighton, BN1 9QH, UK\\
$^{44}$ Cerro Tololo Inter-American Observatory, National Optical Astronomy Observatory, Casilla 603, La Serena, Chile\\
$^{45}$ Brandeis University, Physics Department, 415 South Street, Waltham MA 02453\\
$^{46}$ Instituto de F\'isica Gleb Wataghin, Universidade Estadual de Campinas, 13083-859, Campinas, SP, Brazil\\
$^{47}$ Computer Science and Mathematics Division, Oak Ridge National Laboratory, Oak Ridge, TN 37831\\
$^{48}$ Observatories of the Carnegie Institution for Science, 813 Santa Barbara St., Pasadena, CA 91101, USA\\
\label{lastpage}
\end{document}